\RequirePackage[l2tabu, orthodox]{nag}
\documentclass[
superscriptaddress,
amsfonts,amsmath,amssymb,
reprint,
aps,
pra
]{revtex4-2}

\usepackage[utf8]{inputenc}
\usepackage[german,english]{babel}
\usepackage{graphicx}
\usepackage{float}
\graphicspath{{Figures/}}
\usepackage[dvipsnames,table]{xcolor}
\usepackage[separate-uncertainty=true]{siunitx}
\usepackage{booktabs}
\usepackage[T1]{fontenc}
\usepackage{microtype}
\usepackage{lmodern}
\usepackage{changes}
\usepackage{pdfpages}
\usepackage{afterpage}
\usepackage{braket}
\usepackage{chngcntr}
\usepackage[normalem]{ulem}
\usepackage[colorlinks,
            linkcolor={RoyalBlue},
            citecolor={RoyalBlue},
            urlcolor={RoyalBlue},]{hyperref}
\hypersetup{
    pdftitle={Fabris, Falthansl-Scheinecker et al. "A field-resilient hybrid platform for spin-based quantum devices in planar germanium" (2025)},
    }
\definechangesauthor[color=red]{GK}
\definechangesauthor[color=Green]{GF}
\definechangesauthor[color=blue]{PFS}

\DeclareSIUnit{\sq}{\Box}
\DeclareSIUnit\bar{bar}

\newcommand\gQD{g_{\mathrm{QD}}}
\newcommand\gSC{g_{\mathrm{SC}}}
\newcommand\gMT{g_{\mathrm{MT}}}
\newcommand\geff{g_{\mathrm{eff}}}
\newcommand\GS{\Gamma_\mathrm{S}}
\newcommand\phiLR{\varphi_{\mathrm{LR}}}

\newcommand{\secref}[1]{\hyperref[#1]{{Section~\ref{#1}}}}
\newcommand{\chapref}[1]{\hyperref[#1]{{Chapter~\ref{#1}}}}
\newcommand{\suppref}[1]{\hyperref[#1]{{App.~\ref{#1}}}}
\newcommand{\figref}[1]{\hyperref[#1]{{Fig.~\ref*{#1}}}}
\newcommand{\figrefsupp}[1]{\hyperref[#1]{{Supplementary Fig.~\ref*{#1}}}}
\newcommand{\Figref}[1]{\hyperref[#1]{{Figure~\ref*{#1}}}}
\newcommand{\figrefadd}[2]{\hyperref[#1]{{Fig.~\ref*{#1}#2}}}
\newcommand{\figrefsuppadd}[2]{\hyperref[#1]{{Supplementary Fig.~\ref*{#1}#2}}}
\newcommand{\Figrefadd}[2]{\hyperref[#1]{{Figure~\ref*{#1}#2}}}
\newcommand{\tabref}[1]{\hyperref[#1]{Table~\ref*{#1}}}
\newcommand{\tabrefsupp}[1]{\hyperref[#1]{Supplementary Table~\ref*{#1}}}
\renewcommand{\eqref}[1]{\hyperref[#1]{{Eq.~\ref*{#1}}}}

\makeatletter
\AtBeginDocument{\let\LS@rot\@undefined}
\makeatother

\makeatletter
\newcommand*{\balancecolsandclearpage}{%
  \close@column@grid
  \cleardoublepage
  \twocolumngrid
}
\makeatother

\begin{document}

\title{Granular aluminum induced superconductivity in germanium for hole spin-based hybrid devices}
\author{Giorgio Fabris}
\thanks{These authors contributed equally to this work.}
\affiliation{ISTA,~Institute~of~Science~and~Technology~Austria,~Am~Campus~1,~3400~Klosterneuburg,~Austria}
\author{Paul Falthansl-Scheinecker}
\thanks{These authors contributed equally to this work.}
\affiliation{ISTA,~Institute~of~Science~and~Technology~Austria,~Am~Campus~1,~3400~Klosterneuburg,~Austria}
\author{Devashish Shah}
\affiliation{ISTA,~Institute~of~Science~and~Technology~Austria,~Am~Campus~1,~3400~Klosterneuburg,~Austria}
\author{D. Michel Pino}
\affiliation{Quantum~Advanced~Research~Center~(QuARC),~ Consejo~Superior~de~Investigaciones~Científicas~(CSIC),~Madrid,~Spain}
\affiliation{Instituto de Ciencia de Materiales de Madrid (ICMM),~Consejo Superior de Investigaciones Científicas (CSIC),~Sor Juana Inés de la Cruz 3,~28049 Madrid,~Spain}
\author{Maksim Borovkov}
\affiliation{ISTA,~Institute~of~Science~and~Technology~Austria,~Am~Campus~1,~3400~Klosterneuburg,~Austria}
\author{Anton Bubis }
\affiliation{ISTA,~Institute~of~Science~and~Technology~Austria,~Am~Campus~1,~3400~Klosterneuburg,~Austria}
\author{Kevin Roux}
\affiliation{ISTA,~Institute~of~Science~and~Technology~Austria,~Am~Campus~1,~3400~Klosterneuburg,~Austria}
\author{Dina Sokolova}
\affiliation{ISTA,~Institute~of~Science~and~Technology~Austria,~Am~Campus~1,~3400~Klosterneuburg,~Austria}
\author{Alejandro Andres Juanes}
\affiliation{ISTA,~Institute~of~Science~and~Technology~Austria,~Am~Campus~1,~3400~Klosterneuburg,~Austria}
\author{Tommaso Costanzo}
\affiliation{ISTA,~Institute~of~Science~and~Technology~Austria,~Am~Campus~1,~3400~Klosterneuburg,~Austria}
\author{Inas Taha}
\affiliation{Catalan Institute of Nanoscience and Nanotechnology - ICN2 (CSIC and BIST), Campus UAB, Bellaterra, 08193 Barcelona, Catalonia, Spain}
\author{Aziz Genç}
\affiliation{Catalan Institute of Nanoscience and Nanotechnology - ICN2 (CSIC and BIST), Campus UAB, Bellaterra, 08193 Barcelona, Catalonia, Spain}
\author{Jordi Arbiol}
\affiliation{Catalan Institute of Nanoscience and Nanotechnology - ICN2 (CSIC and BIST), Campus UAB, Bellaterra, 08193 Barcelona, Catalonia, Spain}
\affiliation{ICREA, Pg. Lluís Companys 23, 08010 Barcelona, Catalonia, Spain}
\author{Stefano Calcaterra}
\affiliation{L-NESS, Physics Department, Politecnico di Milano, via Anzani 42, 22100, Como, Italy}
\author{Afonso De Cerdeira Oliveira}
\affiliation{L-NESS, Physics Department, Politecnico di Milano, via Anzani 42, 22100, Como, Italy}
\author{Daniel Chrastina}
\affiliation{L-NESS, Physics Department, Politecnico di Milano, via Anzani 42, 22100, Como, Italy}
\author{Giovanni Isella}
\affiliation{L-NESS, Physics Department, Politecnico di Milano, via Anzani 42, 22100, Como, Italy}
\author{Rub\'en Seoane Souto}
\affiliation{Quantum~Advanced~Research~Center~(QuARC),~ Consejo~Superior~de~Investigaciones~Científicas~(CSIC),~Madrid,~Spain}
\affiliation{Instituto de Ciencia de Materiales de Madrid (ICMM),~Consejo Superior de Investigaciones Científicas (CSIC),~Sor Juana Inés de la Cruz 3,~28049 Madrid,~Spain}
\author{ Maria Jos\'e Calder\'on}
\affiliation{Quantum~Advanced~Research~Center~(QuARC),~ Consejo~Superior~de~Investigaciones~Científicas~(CSIC),~Madrid,~Spain}
\affiliation{Instituto de Ciencia de Materiales de Madrid (ICMM),~Consejo Superior de Investigaciones Científicas (CSIC),~Sor Juana Inés de la Cruz 3,~28049 Madrid,~Spain}
\author{Ram\'on Aguado}
\email{ramon.aguado@csic.es}
\affiliation{Quantum~Advanced~Research~Center~(QuARC),~ Consejo~Superior~de~Investigaciones~Científicas~(CSIC),~Madrid,~Spain}
\affiliation{Instituto de Ciencia de Materiales de Madrid (ICMM),~Consejo Superior de Investigaciones Científicas (CSIC),~Sor Juana Inés de la Cruz 3,~28049 Madrid,~Spain}
\author{Jos\'e Carlos Abadillo-Uriel}
\email{jc.abadillo.uriel@csic.es}
\affiliation{Quantum~Advanced~Research~Center~(QuARC),~ Consejo~Superior~de~Investigaciones~Científicas~(CSIC),~Madrid,~Spain}
\affiliation{Instituto de Ciencia de Materiales de Madrid (ICMM),~Consejo Superior de Investigaciones Científicas (CSIC),~Sor Juana Inés de la Cruz 3,~28049 Madrid,~Spain}
\author{Georgios Katsaros}
\email{georgios.katsaros@ist.ac.at}
\affiliation{ISTA,~Institute~of~Science~and~Technology~Austria,~Am~Campus~1,~3400~Klosterneuburg,~Austria}

\begin{abstract}

In superconductor-semiconductor hybrid structures superconductivity and spin-polarization are competing effects as magnetic fields break Cooper pairs. They can be combined using thin films and in-plane magnetic fields, an approach that enabled the pursuit of Majorana zero modes, Kitaev chains, and Andreev spin qubits (ASQs), but remains challenging for materials with small in-plane $g$-factors or when out-of-plane fields are required. We demonstrate that granular aluminium (grAl), composed of nanometre-scale aluminium grains embedded in an amorphous oxide matrix, can overcome this limitation. By depositing grAl on Ge/SiGe heterostructures, we induce a hard superconducting gap with BCS peaks at 305 $\mu$eV and magnetic field resilience for both the in-plane and out-of-plane direction allowing Zeeman splitting of Yu–Shiba–Rusinov (YSR) states beyond 50 $\mu$eV (12 GHz).  Leveraging this robustness, we reveal signatures of hole physics and discuss a driving mechanism for ASQs regardless of the strength of the Rashba spin-orbit coupling in planar Germanium.

\end{abstract}

\maketitle

\section*{Introduction}
\label{sec:intro}

Silicon (Si) and germanium (Ge) are suitable materials for building scalable quantum processors owing to their promising spin qubit properties \cite{RevModPhys.95.025003,Scappucci2021}. 
Achieving superconductivity in such group-IV materials would further open the path to quantum dot (QD) based hybrid superconducting-semiconducting devices such as Kitaev chains \cite{dvir2023realization, ten2024two}, Cooper pair splitters (CPSs) \cite{hofstetter2009cooper, strunk2010, Bordoloi2022, wang2022singlet, wang2023triplet} and ASQs \cite{Hays2021ASQ, pita2023direct, hoffman2025resolvingandreevspinqubits}. 
However, superconductivity in Si has been achieved only with high boron doping levels slightly below~\cite{oldSi} or above~\cite{bustarret2006superconductivity,chiodi2017proximity} the solubility limit, an approach that is rather difficult to adapt for nanoscale devices. On the other hand, in Ge, superconductivity has been achieved not only through doping \cite{Steele2025}, but also via the superconducting proximity effect~\cite{reviewnovelhybrid}. Multiple theoretical descriptions
\cite{adelsberger2023microscopic,pino2025theory,babkin2025superconducting} and experimental realizations have been reported, ranging from nanowires and self-assembled nanocrystals to strained Ge quantum wells (QWs) \cite{Xiang2006,katsaros2010hybrid, de2018spin,vigneau2019germanium,Ridderbos2018,aggarwal2021enhancement,hendrickx2018gate, tosato2023hard, hinderling2024direct, valentini2024parity, Leblanc2025}, the most suitable for spin applications.
However, in the latter the $g$-factor is relatively small for in-plane fields (around 0.2-0.5 \cite{jirovec2022dynamics, hendrickx2024sweet, seidler2025spatial, Miller2022}), which makes it challenging to achieve spin polarization before superconductivity breaks down. Moreover, even if the superconductor could withstand the required magnetic field, spin coherence would remain limited, since the dephasing time for holes in Ge has been reported to scale inversely with the magnetic field \cite{lawriePhD}. Being able to use out of-plane fields in isotopically purified heterostructures would not only reduce the required field magnitude due to the much larger $g$-factors but also make use of the aligned spin-quantization axis \cite{jirovec2022dynamics, hendrickx2024sweet, seidler2025spatial, saez2025exchange}, favorable for Pauli spin blockade readout. 

Establishing magnetic-field compatibility in proximitized QDs is therefore a key ingredient for the realization of ASQs, Kitaev chains and CPSs. In addition, having a large superconducting gap would benefit these hybrid quantum devices through reduced quasiparticle poisoning. For Kitaev chains specifically, increased spectral separation prevents hybridization between zero-energy Majorana modes and the quasiparticle continuum, thereby maintaining topological protection.

From the above, it becomes evident that a superconductor inducing a large hard gap in Ge with resilience in all magnetic field directions would unlock the potential of planar Ge for spinful hybrid devices. So far a hard gap has been achieved in \cite{tosato2023hard} with PtSiGe superconducting contacts and in \cite{valentini2024parity} using Al as the parent superconductor, with gap sizes of 71 $\mu$eV and 150 $\mu$eV (coherence peaks position), respectively. Both these approaches have limited out-of-plane resilience, which can be enhanced by realizing leads with widths down to tens of nanometers as done in \cite{lakic2025quantum} to realize a proximitized QD. 
Here, we opted for a superconductor which has proven to give a hard gap in Ge, namely Al, and we introduced controlled disorder to boost the magnetic field resilience in all directions even for rather wide ($\sim\mu$m) and thick films ($\sim20$nm).
Our findings establish that grAl, a superconductor composed of nanoscale Al grains embedded in an oxide matrix \cite{cohen1968superconductivity, Deutscher1973, Levy-Bertrand2019, moshe2020}, offers a unique combination of properties for advancing Ge-based quantum technologies. While its enhanced kinetic inductance and magnetic field resilience have already been discussed as advantageous for high-impedance circuit applications \cite{maleeva2018circuit, grunhaupt2019granular, Rotzinger_2017}, we uncover a second critical capability. grAl induces a hard superconducting gap with BCS coherence peaks up to 305 $\mu$eV in a Ge QW.
This large gap, combined with its kinetic inductance and magnetic-field resilience, establishes grAl as an versatile platform for building Ge-based quantum processors. We use this platform to proximitize a gate-defined QD and investigate the magnetic field response of a YSR state at different QD-superconductor coupling strengths, revealing two signatures of the hole-like nature of the confined carriers: a tunable anisotropic g-tensor and the spin-dependent corrections to the tunnel couplings which are magnetic field-dependent. In what follows we dub the latter effect magnetotunneling following~\cite{rodriguez2025sweet} and we further demonstrate theoretically that it will allow the driving and reading out of ASQs regardless of the strength of the Rashba spin-orbit coupling.

\section*{Results}
\subsection*{GrAl induced proximity effect}
\label{sec:grAl}

\begin{figure*}[htbp!]
    \centering
    \includegraphics{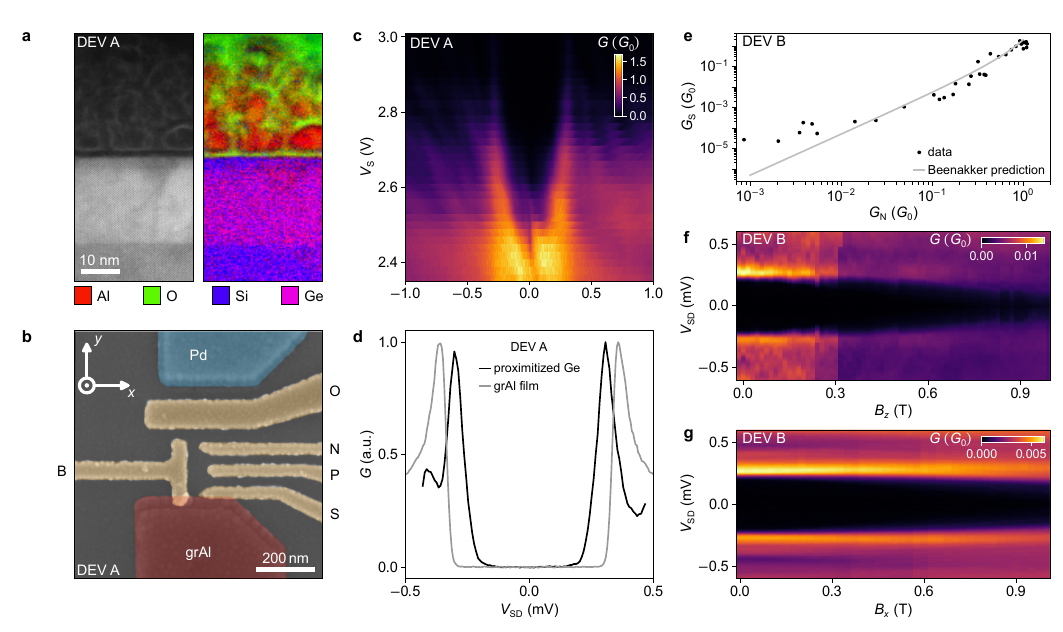}
    \caption{\textbf{GrAl induced superconductivity in planar Ge.} \textbf{a.} High-angle annular dark-field scanning transmission electron microscopy image of a Ge/SiGe heterostructure with a grAl layer on top (left), and the corresponding electron energy-loss spectroscopy (right). The granular structure of the grAl layer is clearly resolved. A thin oxide layer is observed at the interface between the grAl layer and the QW. In addition, a localized Al region is detected within the QW 
    (see  SI Fig. S1).
    \textbf{b.} False-colour scanning electron microscopy image of a copy of DEV A, with the superconducting grAl contact in red, the Pd contact in light-blue and gates in yellow. B and S are used to electrostatically form a tunable constriction to probe the proximitized region. Together with N and P they are used to form a QD, where N controls the coupling to the normal lead and P the QD electrochemical potential. Gate O tunes the Ge density of states close to the dot. For all the results presented in the main text, it was grounded. \textbf{c.} DEV A differential conductance $G$ = $dI/dV$ in units of $G_0 = 2e^2/h$ as a function of the voltage applied to gate S ($V_{\mathrm{S}}$) and bias voltage ($V_{\mathrm{SD}}$) at $V_\mathrm{B}$ = 4.35 V, with gates $V_{\mathrm{P}}$ and $V_{\mathrm{N}}$ set at 0V so that no QD is formed.  At high $V_{\mathrm{S}}$ values (tunneling regime) coherence peaks at $\pm$ 305 $\mu$V are observed. Around $V_{\mathrm{S}}$ = 2.4 V (single channel regime), we measure enhanced in-gap conductance approaching 2$\mathrm{G_0}$ (see SI Fig. S3).
    \textbf{d.} Normalized $G$ vs $V_{\mathrm{SD}}$ for proximitized Ge (black line) in DEV A and for grAl (grey line). The black trace was obtained at $V_\mathrm{S}$ = 3.26 V, deep in the tunneling regime, where the peak conductance is $6\times10^{-3}G_0$. The grey trace shows the numerical derivative $dI/dV_\mathrm{SD}$ of the current measured in tunneling spectroscopy of a grAl film with the same resistivity as that used in this study (see methods).
    For proximitized Ge the BCS peak is at $\Delta^*$ = 305 $\mu$V, to be compared with $\Delta_{\mathrm{grAl}}$ = 360 $\mu$V. \textbf{e.} Zero-bias conductance in the superconducting state ($G_S$) plotted against the corresponding normal-state value ($G_N$) for DEV B. The normal-state conductance ($G_N$) is obtained by locally heating the chip above the grAl critical temperature using an on-chip resistor. After the device cools back down, the superconducting conductance ($G_S$) is measured under identical gate conditions. Each data point corresponds to a $V_\mathrm{S}$ value, while $V_\mathrm{B}$ is kept fixed. Data points match well with the theoretical prediction $G_\mathrm{S} = 2G_0\frac{G_\mathrm{N}^2}{(2G_0-G_\mathrm{N})^2}$. \textbf{f-g.} $G$ vs $V_{\mathrm{SD}}$ and magnetic field $B_{z,x}$ of DEV B, where the grAl contact is 300 nm wide and 20 nm thick. A region of reduced subgap conductance is visible up to around 1 T in both directions.}
    
    \label{fig:fig1}
\end{figure*}

\begin{figure*}[htbp!]
    \centering
    \includegraphics{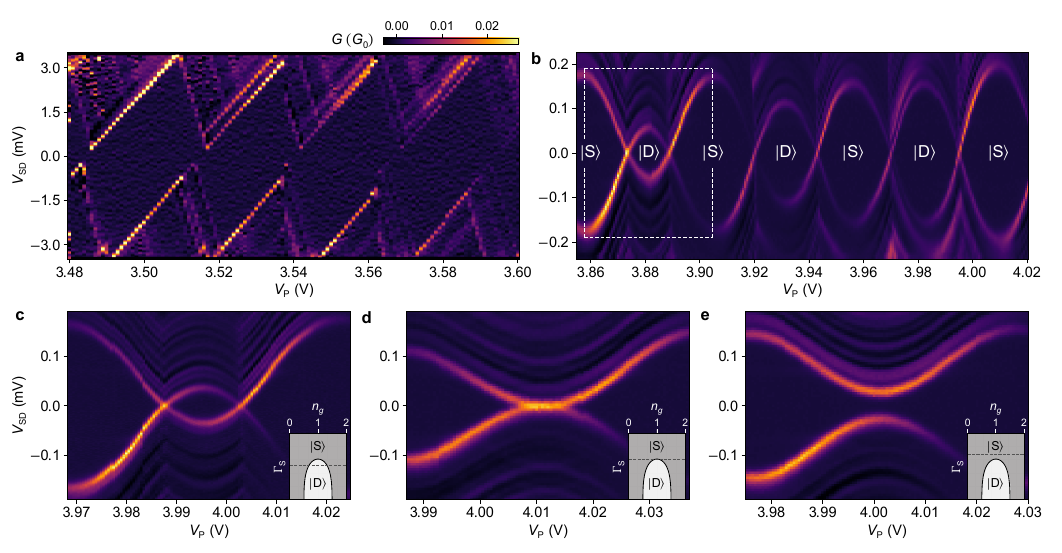}
    \caption{\textbf{Spectroscopy measurements of a hybrid QD for different couplings to the supercondcutor in DEV A.}
    \textbf{a.} $G$ through the QD vs $V_\mathrm{P}$ and $V_{\mathrm{SD}}$ calculated by numerical differentiation of the measured current (G = $dI/dV_\mathrm{SD}$). Coulomb diamonds with a region of suppressed conductance for $|V_{\mathrm{SD}}|$ < 200 $\mu$V and a charging energy of $U$ = 3.2 mV are observed. \textbf{b.} $G$, measured with a lock-in amplifier, at increased coupling to the superconductor (lower $V_\mathrm{S}$) compared to the configuration shown in (a). Subgap states emerge displaying the characteristic singlet-doublet ground state sequence associated with YSR physics. At higher bias voltage, additional conductance features that replicate the YSR state appear. Their origin is discussed in the SI Fig. S4. 
    \textbf{c-e.} $G$ vs $V_\mathrm{P}$ and $V_{\mathrm{SD}}$ with increasing $\Gamma_{\mathrm{S}}$ for the YSR enclosed by the white dashed line in (b). 
   For a fixed $\Gamma_\mathrm{S}$ sweeping $V_\mathrm{P}$ corresponds to sweeping the effective dot occupancy $n_g$ in the singlet-doublet phase diagram (see inset), where $n_g=1$ corresponds to the center of the doublet region. At moderate coupling strengths the system exhibits well-defined singlet and doublet regions, while beyond a critical $\Gamma_\mathrm{S}$ (d) a singlet-like ground state persists across the full $n_g$ range (e). For the ZBW model we extract $U =$ 685 $\mu$eV and $\Gamma_\mathrm{S}$ = 156 $\mu$eV (c), 183 $\mu$eV (d), 204 $\mu$eV (e) (see SI section S6).}
    \label{fig:fig2}
\end{figure*}

\begin{figure*}[htbp!]
    \centering
    \includegraphics{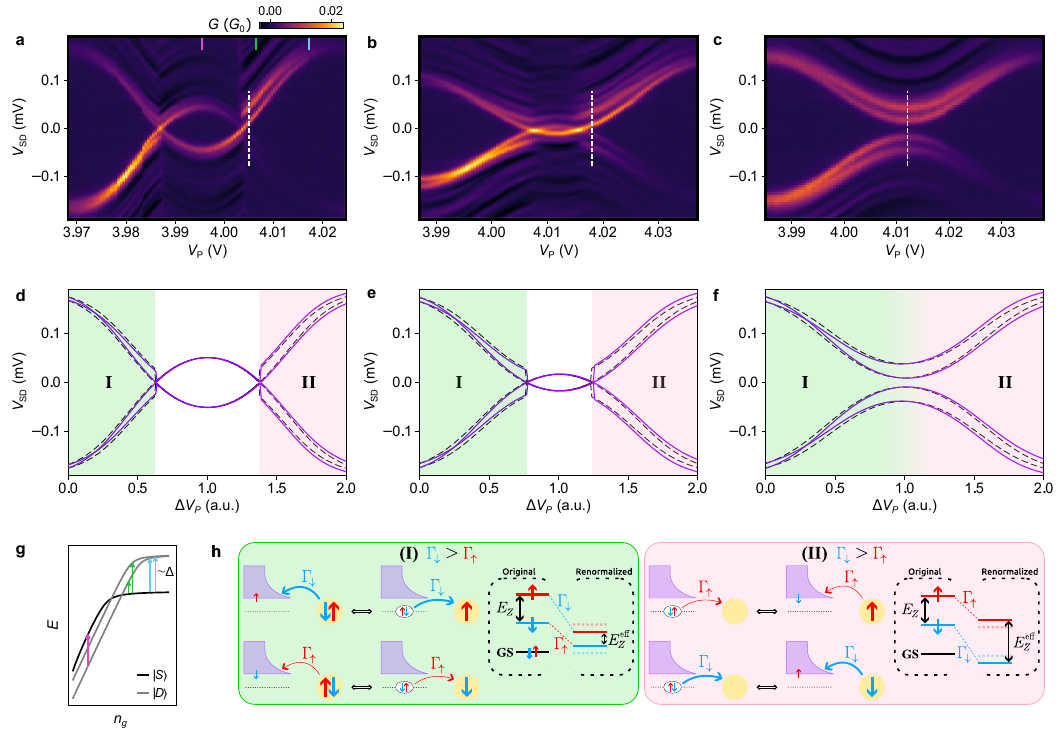}
    \\
    \caption{\textbf{Spin splitting of YSRs revealing signatures of hole physics in DEV A.}
    \textbf{a-c.} $G$ vs $V_\mathrm{P}$ and $V_{\mathrm{SD}}$ in the same coupling regimes as \figrefadd{fig:fig2}{c-e}, respectively, with an applied out-of-plane magnetic field of 80 mT. Pronounced splitting is observed in regions with a singlet ground state.
    In (a-b) the Zeeman splitting is asymmetric between the left and right sides of the doublet ground-state region. The white dashed lines indicate the plunger voltages for measurements in \figrefadd{fig:fig4}{}. Additional data showing similar asymmetry can be found in section S6 of the SI. \textbf{d-f} Lowest-energy even-odd parity transitions calculated with the ZBW model. Solid lines include the magnetotunneling effect due to heavy-hole-light-hole mixing. Dashed lines show the result without magnetotunneling for comparison. We use $\Gamma_\mathrm{S}=156, \,183 \; {\rm and} \; 204 \; \mu$eV as extracted from \figrefadd{fig:fig2}{c-e}, respectively. \textbf{g} Lowest energy levels of the system as a function of
    $n_g$, where only a section of the energy spectrum is shown, corresponding to the right-half of (a). The cartoon is obtained using a ZBW model without the magnetotunneling term and without the Zeeman term in the superconductor Hamiltonian. The gray lines correspond to the Zeeman-split lowest-energy doublet state, whereas the black line correspond to the lowest-energy singlet state. The colored arrows indicate the transport-allowed transitions at different values of $n_g$ corresponding to the colored lines in panel a. \textbf{h.} Cartoon illustrating the lowest-order tunneling processes between the superconductor and the QD in regions I and II. On each region, the $\Leftrightarrow$ symbol indicates the relevant tunneling transitions between the QD and superconductor. The thick blue and thin red arrows denote processes where a spin down and a spin up electron is exchanged between the QD and the superconductor.  The thickness of each line illustrates the amplitude of each process, $\Gamma_\downarrow>\Gamma_\uparrow$. The right-side insets show the resulting effect of the hybridization on the spectrum. Since $\Gamma_\downarrow\neq\Gamma_\uparrow$ and the spin-tunneling processes on each region are the opposite, the Zeeman splitting $E_Z$ is renormalized to $E_Z^\text{eff}$ asymmetrically for each region.}
    \label{fig:fig3}
\end{figure*}

\begin{figure*}[htbp!]
    \centering
    \includegraphics{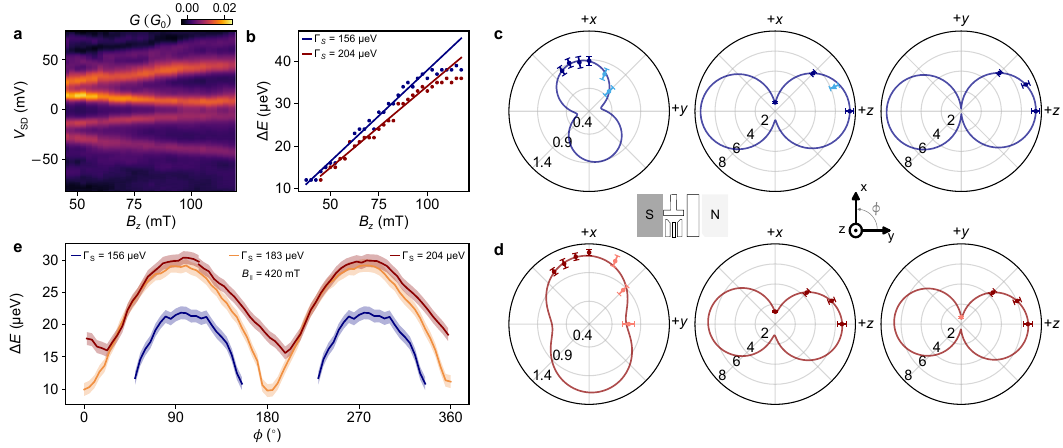}
    \caption{\textbf{$g$-tensor measurements for different coupling strengths in DEV A.} \textbf{a.} $G$ vs $V_{\mathrm{SD}}$ and $B_z$ (out-of-plane direction) at $\Gamma_\mathrm{S}$ = 204 $\mu$eV and with $V_\mathrm{P}$ fixed at the positions marked in \figrefadd{fig:fig3}{c} by the white dashed line. The splitting initially evolves linearly according to the Zeeman effect. Above roughly 100 mT the splitting deviates from a linear trend, likely due to the interaction with the quasiparticles continuum, as in \cite{lee2014spin}, or to orbital effects. \textbf{b.} Plot showing the Zeeman splitting as a function of $B_z$ for configurations $\square$ ($\Gamma_\mathrm{S}$ = 156 $\mu$eV) and $\bigcirc$ ($\Gamma_\mathrm{S}$ = 204 $\mu$eV). The data points are extracted from the separation of the conductance peaks at positive bias. Solid lines are the results of the linear fits to the data below 105 mT, yielding a $g$-factor of 7.44$\pm$0.20 for $\square$ and of 6.65$\pm$0.23 for configuration $\bigcirc$. \textbf{c,d.} Polar projections on the laboratory frame of the $g$-tensors for configurations \figrefadd{fig:fig3}{a,c} respectively. 
     Dark data points denote linear fits of the Zeeman splitting vs magnetic field magnitude at a fixed direction. Light data points provide a consistency check (see SI section S10). \textbf{e.} Zeeman splitting $\Delta E$ as a function of in-plane magnetic field angle at $B_\parallel$ = 420 mT, where $\phi$ = 0 corresponds to $y$-direction, for the three configurations reported in \figrefadd{fig:fig3}{a-c}. The shaded regions correspond to the standard deviation on the measured splittings, which we consider to be equal to twice the bias voltage resolution (1 $\mu$V).
     }
    \label{fig:fig4}
\end{figure*}
 
The evaporation of Al in an oxygen atmosphere stands in contrast to the past decade’s efforts to achieve ultraclean superconductor-semiconductor interfaces by high-vacuum, low-temperature, in-situ deposition of Al \cite{Chang2015, kjaergaard2016qpc, shabani2016, Moehle2021}. Despite the counterintuitive approach, we demonstrate that grAl can effectively induce superconductivity in a Ge two-dimensional hole gas (2DHG) by straightforward room-temperature deposition on top of a Ge/SiGe heterostructure with a 4 nm $\text{Si}_{0.3}\text{Ge}_{0.7}$ spacer, yielding the cross-section shown in \figrefadd{fig:fig1}{a}.
To assess the induced gap properties, we investigate electronic transport across the interface between the normal and proximitized Ge regions in two different devices (DEV A and DEV B) having similar geometries. Gates S and B (\figrefadd{fig:fig1}{b}) define a tunable constriction in the 2DHG, enabling access to the single-channel regime with continuously adjustable transparency. The transport properties of such constrictions are well understood \cite{Beenakker1992, Cuevas1996}, with the limiting cases providing clear physical intuition. For unity transmission, an electron incident from the normal side is perfectly Andreev reflected as a hole while a Cooper pair is transferred into the superconductor, enhancing the zero-bias conductance to twice the conductance quantum, 2$G_0$~\cite{kjaergaard2016qpc}. In DEV A, this regime is realized at the lowest gate voltages Vs (\figrefadd{fig:fig1}{c}), where the subgap conductance approaches 1.9$G_0$ [see also supplementary information (SI), Fig. S3] with a small dip reaching 1.35$G_0$ at zero bias, which we attribute to normal scattering at the NS boundary \cite{Cuevas1996}. In the tunneling limit (higher Vs), Andreev processes are weak, resulting in suppressed subgap conductance. In this regime, the superconducting gap manifests as coherence peaks in the differential conductance at $\pm$ 305 $\mu$V, corresponding to about 85$\%$ of the gap of granular aluminum (grAl) at the same resistivity (\figrefadd{fig:fig1}{d}). We measure the induced gap to be hard, with subgap conductance around two orders of magnitude lower than out-of-gap conductance \cite{Chang2015} for $|V_\mathrm{SD}| < 180 \mu$eV (see SI sections S2 and S3 for further details). In this device, upon applying magnetic field, we measure conductance below $\mathrm{6}\times\mathrm{10^{-5}}$ $G_0$ (noise level) at zero bias up to 160 mT in the out-of-plane direction and 800 mT in-plane
for a 500 nm wide and 20 nm thick grAl contact (see SI section S2). Similar results have been obtained in multiple devices, with contacts up to 3 $\mu$m wide (SI section S2). To further improve the magnetic field resilience we have realized DEV B, where grAl resistivity is three times higher than in DEV A and the contact width is 300 nm. At this resistivity we still induce a hard gap (\figrefadd{fig:fig1}{e}) with zero bias conductance below $\mathrm{6}\times\mathrm{10^{-5}}$ $G_0$ up to 650 mT in the out-of of plane direction and 1 T in-plane (\figrefadd{fig:fig1}{f,g}).

\subsection*{Hybrid dot}
\label{sec:hybrid_dot}

Next, we employ DEV A to proximitize a QD by engaging the plunger and the normal barrier gates (P and N in \figrefadd{fig:fig1}{b}). The physics of a QD coupled to a superconductor is governed by the competition between superconducting pairing, $\Delta$, and the charging energy, $U$ \cite{reviewnovelhybrid,review2011josephsontransport}. While the superconductor favours even-parity singlet ground states $\ket{S}$, Coulomb repulsion can favour odd occupancy of the QD (doublet ground state $\ket{D}$).
Depending on the relative magnitudes of $U$ and $\Delta$, the singlet state manifests either as an Andreev bound state ($U$<$\Delta$) or as a YSR state ($U$>$\Delta$), the latter arising from the screening of the dot’s unpaired spin by quasiparticles in the superconductor. In the regime of weak coupling to the superconductor, we observe Coulomb diamonds gapped by twice $\Delta$, consistent with the suppression of sequential single-particle transport (\figrefadd{fig:fig2}{a}). From these measurements, we extract a charging energy of around 3 mV, approximately ten times larger than $\Delta$.
Upon increasing the coupling to the superconductor $\Gamma_\mathrm{S}$ by lowering $V_\mathrm{S}$, in-gap states emerge, exhibiting the characteristic eye-shaped dispersion of YSR states (\figrefadd{fig:fig2}{b}) \cite{pillet2011tunn, jellinggaard2016tuning}.  Focusing on a single orbital manifold, we validate this interpretation by fitting the data with a zero-bandwidth (ZBW) model (see SI section S6), from which we extract the ratio $U$/$\Delta$ = 3.5, confirming the YSR limit. We then track the evolution of this state with increasing $\Gamma_\mathrm{S}$, observing a transition from alternating ground-state parity to a singlet-like ground state over the full plunger range (\figrefadd{fig:fig2}{c-e}), in agreement with theoretical expectations. For intermediate coupling (\figrefadd{fig:fig2}{d}), the singlet and doublet states are degenerate within measurement accuracy, around the offset charge $n_g =1$. 

\subsection*{Spin-split YSRs}
\label{sec:Spin-split YSRs}

 By applying an out-of-plane magnetic field of 80 mT, we observe Zeeman splitting exceeding the thermal broadening, allowing spin polarization of the QD without compromising superconductivity (\figrefadd{fig:fig3}{a-c}). As expected and shown in \figrefadd{fig:fig3}{g}, pronounced splitting is observed in regions with a singlet ground state.
 When the singlet and doublet states are degenerate and a magnetic field is applied, the system transitions into a doublet ground state, marked by the emergence of the characteristic eye-shaped dispersion (\figrefadd{fig:fig3}{b}).
For all three hybridization strengths, the splitting decreases as the system is tuned deeper into the singlet ground state regime. We attribute this behaviour to level repulsion from the continuum of quasiparticle states in the superconductor, consistent with previous works \cite{lee2014spin,jellinggaard2016tuning}. We note that the ZBW description is therefore most reliable near the singlet–doublet boundary and progressively underestimates the YSR renormalization deeper in the singlet regions, where hybridization with the continuum plays a stronger role.

Strikingly, the spin splitting is not symmetric on the two sides of the eye-shaped ground-state doublet region (pink mark in~\figrefadd{fig:fig3}{a}).
For example, this can be clearly seen in~\figrefadd{fig:fig3}{a} where the spin-splitting is much larger on the right side of the lobe than on the left side, where a level crossing is observed near $V_\text{P}\approx3.98$ V. Similar behavior is found for larger hybridization strengths (\figrefadd{fig:fig3}{b}) with  the level crossing around $V_\text{P}\approx4.00$ V. 
Translating these observations to $g$-factors results in a dramatic change over a small plunger voltage range of 20 mV. This corresponds to an effective $g$-factor evolving from nearly zero to $g\approx 7.44$ in~\figrefadd{fig:fig3}{a}, as discussed in the next section. While gate-tuneable $g$-factors are well established for hole spins, reported changes are typically $dg/dV_{P}$ $\approx 10^{-3}\mathrm{mV}^{-1}$~\cite{piot2022single, hendrickx2024sweet}, orders of magnitude smaller than what we observe.

This drastic change in the spin-splitting with gate voltage requires a more detailed analysis. In~\figrefadd{fig:fig3}{d-f}, we show that a minimal extension of the ZBW model reproduces the pronounced spin-splitting change with gate voltage. The only added ingredient is a magnetic correction to the QD - superconductor coupling strength such that $\Gamma_{\uparrow/\downarrow}=\Gamma_\mathrm{S}\pm\Gamma_\text{MT}(\mathbf{B})$
(details in methods and SI section S5). This \emph{magnetotunneling} effect arises from heavy-hole–light-hole admixture and was recently identified as a strong $g$-factor renormalization mechanism in hole-based coupled double QDs~\cite{rodriguez2025sweet}. We take $\Gamma_\text{MT}=0.06\Gamma_\mathrm{S}$ for the magnetic field used in the measurements in \figrefadd{fig:fig3}{a-c}, not only reproducing the asymmetry but also the observed crossing between the doublet states. For clarity, we distinguish the left and right singlet-ground-state regions in these plots as regions I and II, respectively. The sign of the correction $\Gamma_\text{MT}(\mathbf{B})$ determines which region has larger or smaller splitting, as shown in the Methods section. 

In Figure~\figrefadd{fig:fig3}{h} we provide a more intuitive picture of the asymmetry between the two regions. In this figure, we emphasize that the measured states are not pure QD states but hybrid QD-superconductor states whose energies depend on how strongly the QD spin states hybridize with the superconductor. On both regions, the ground state is a singlet, but the odd excitation is built in two different ways depending on the gate setting. In region~I, the lowest odd excitation is obtained by taking one hole out of the QD singlet, whereas in region~II it is obtained by putting one hole into the QD singlet. Magnetotunneling makes this hybridization spin dependent ($\Gamma_\uparrow\neq\Gamma_\downarrow$). As a consequence, the two spin-polarized odd excitations couple with different strength to the superconductor and acquire different energy shifts. Each region in panel~\figrefadd{fig:fig3}{h} depicts one such tunneling process, with the double-headed arrow $\Leftrightarrow$ connecting the two configurations linked by tunneling events at rate $\Gamma_\uparrow$ or $\Gamma_\downarrow$ (taking $\Gamma_\downarrow>\Gamma_\uparrow$). In region~I, the relevant rates are $\Gamma_\downarrow$ (top row, controlling the $\uparrow$ branch) and $\Gamma_\uparrow$ (bottom row, controlling the $\downarrow$ branch); in region~II, where the process is inverted, the assignment of $\Gamma_\uparrow$ and $\Gamma_\downarrow$ to the two spin branches is exchanged. This interchange naturally produces the observed left-right asymmetry of the spin splitting. The right-side insets translate this into the spectrum: starting from the bare Zeeman-split QD levels $E_Z$, the spin-dependent coupling shifts the up and down branches by different amounts, yielding the renormalized level splitting $E_Z^\mathrm{eff}$. Dashed lines mark the level positions in the absence of magnetotunneling ($\Gamma_\uparrow=\Gamma_\downarrow$) and serve as a reference. See Methods for the minimal model and SI section~S5 for a more detailed derivation.

\subsection*{$g$-tensors}
\label{sec:g-tensors}

In addition to the magnetotunneling contribution, coupling to the superconductor is itself expected to renormalize the QD $g$-tensor. Specifically, holes in planar Ge are known to exhibit a strongly anisotropic $g$-tensor, with the $g$-factor ranging from $\sim$7 in the out-of-plane direction to $\sim$0.3 in-plane. Hybridization of a QD with a superconductor characterized by an isotropic $g$-factor of 2 can lead to a reduction of the out-of-plane $g$-factor and an enhancement of the in-plane components \cite{babkin2025superconducting}. Leveraging the field robustness of the grAl contact and the stability of the device, we measured in transport the full $g$-tensor of the YSR under study for the two different coupling strengths used in \figrefadd{fig:fig2}{c} and \figrefadd{fig:fig2}{e}. Field sweeps have been taken at positions indicated by the white dashed lines in \figrefadd{fig:fig3}{a,c} (see \figrefadd{fig:fig4}{a} for example field sweep) and the extracted Zeeman splittings have been fitted to a line to extract the corresponding $g$-factors (\figrefadd{fig:fig4}{b}). Figures \figrefadd{fig:fig4}{c,d} show the resulting $g$-tensors. We observe larger in-plane $g$-factors than commonly reported. While previous studies typically report in-plane values of around 0.2-0.5, here we demonstrate a maximum of $g$=1.25 for the strongest hybridization, effectively halving the in-plane magnetic fields required for spin-related applications. Furthermore , we show an enhancement of the in-plane $g$-tensor projection and a reduction of the out-of-plane component with increasing $\Gamma_\mathrm{S}$, consistent with the renormalization hypothesis.

\begin{figure*}[htbp!]
    \centering
    \includegraphics{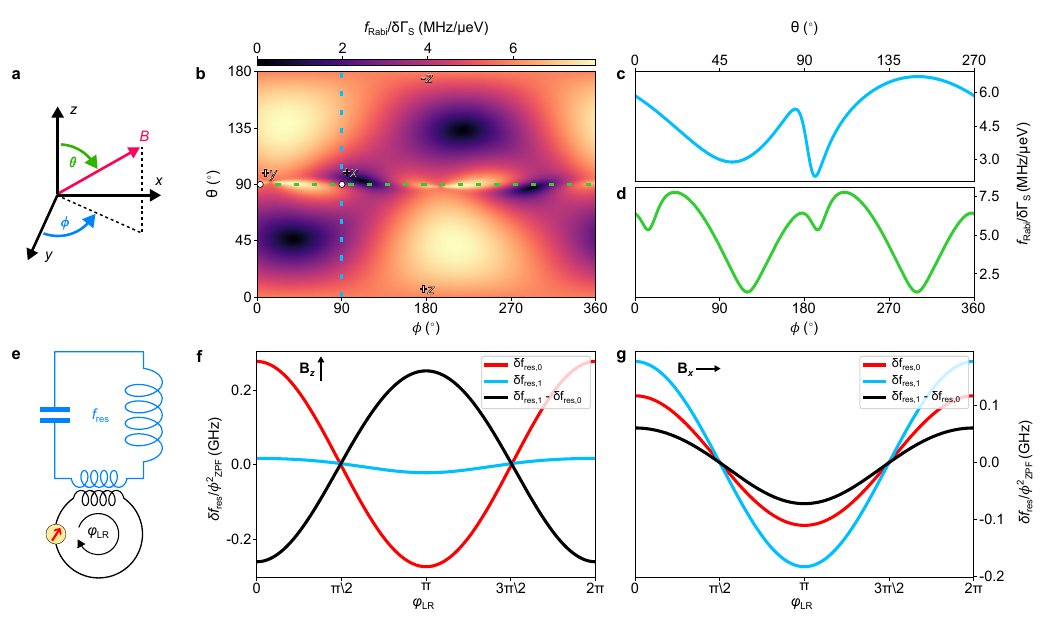}
    \caption{\textbf{Hybridization-driven spin control and phase-sensitive resonator detection.}
    \textbf{a.} Magnetic field-angle convention: $\theta$ from $+\hat{z}$, $\phi=0$ along $+\hat{y}$ increasing towards $+\hat{x}$. (b) $g$-tensor-modulation Rabi frequency per unit modulation of the QD-superconductor coupling, $f_\mathrm{R}/\delta\GS$ ($\delta\GS$ as the amplitude of the sinusoidal driving), versus magnetic-field orientation at $|\mathbf{B}|=100$~mT, $\GS^0=156~\mu$eV and $n_g=0.58$ (region II). \textbf{c,d.} Cuts along the blue (vertical) and green (horizontal) dashed lines in \textbf{b}. \textbf{e.} Phase-biased SC-QD-SC junction coupled to a resonator at $n_g=1$ where the local one-lead magnetotunneling correction vanishes. In this geometry, virtual transitions through additional off-resonant levels are expected to contribute a finite cotunneling $t_c$ between the superconductors, as in electron-based Andreev spin qubits~\cite{BargerbosPRL2023}. \textbf{f,g.} State-dependent adiabatic resonator curvature $\delta f_{\mathrm{res},s}/\varphi_\mathrm{ZPF}^2=h^{-1}\partial^2E_s/\partial\phiLR^2$ in the geometry of \textbf{e} with relative phase $\phiLR$ between superconductors, for $\mathbf{B}\parallel z$ and $\mathbf{B}\parallel x$ at $|\mathbf{B}|=100$~mT. Red and blue are the two Zeeman branches $E_s=E_D(\phiLR)+s\frac{\mu_B}{2}|\geff(\phiLR)\mathbf{B}|$, labelled state 0 ($s=+1$) and state 1 ($s=-1$); black is their difference, the readout-relevant signal. Parameters in SI section S14.}
    \label{fig:outlook_control_readout}
\end{figure*}

We stress, that the two coupling regimes are not probed under exactly equivalent conditions. For strong QD-superconductor hybridization, the spin splitting is extracted close to the symmetric point ($n_g=1$), where the magnetotunneling contribution vanishes. For weaker coupling, the symmetric point belongs to the doublet ground-state region, so the splitting must be measured at a different gate voltage, where magnetotunneling contributes to the $g$-tensor. 
Even with this limitation, the dominant trend is consistent with QD-superconductor hybridization. In particular, if the enhancement of the out-of-plane $g$-factor at lower coupling were mainly due to magnetotunneling, a similar enhancement would be expected for in-plane magnetic fields since the asymmetry is similar for this configuration (see Fig. ~S6 for in-plane YSR spectroscopy). Instead, the in-plane splittings increase with increasing QD-superconductor coupling (see \figrefadd{fig:fig4}{e}), which argues against a magnetotunneling-dominated interpretation. We therefore attribute the main change of the measured $g$-factors to hybridization with the superconductor, with magnetotunneling and gate-dependent $g$-factor corrections acting as secondary corrections.

Finally, we mention that by further reducing the coupling to the superconductor, we measured a strong reduction of the $g$-factor in both the in- and out-of-plane directions (Fig. S11 in SI), which might be due to variations in the orbital wavefuctions induced by changes in gate voltages.

\section*{Discussion/Outlook}
\label{sec:discussion}

In this work, we have established a novel platform for spin-based hybrid devices in planar Ge. We demonstrated this by extracting the full $g$-tensor of a  QD-defined YSR state, revealing an in-plane $g$-factor of up to 1.25 and out-of-plane $g$-factors between 6.65 and 7.44, depending on the coupling strength to the superconductor.
With these $g$-factors, achieving Zeeman splittings above thermal broadening and in the typical microwave range used for spin-qubit control and circuit quantum electrodynamics (cQED) architectures (e.g., 5 GHz) requires an in-plane magnetic field of 360 mT or an out-of-plane field of 50 mT. The proximitized Ge remains resilient at such fields, showing no degradation of the induced YSR states. This enabled us to observe a pronounced asymmetry in the Zeeman splitting on the two sides of the doublet ground state region, which we attribute to heavy-hole-light-hole mixing. A more quantitative understanding of this phenomenon will require operating at lower hole occupancies, which is out of the scope of the present work.

Apart from correcting the spectrum, the field-induced spin-dependent correction to tunneling can provide a useful dynamical resource. Modulating $\GS$ changes the relative QD, superconductor, and magnetotunneling contributions to the total effective Zeeman field. Specifically, in the most general case for an arbitrary magnetic field the tunneling becomes $\GS\rightarrow \GS\mathbb{I}-\frac{\mu_B}{2}\boldsymbol{\sigma}\cdot \gMT(\GS)\mathbf{B} $ which reduces to the scalar $\GS\pm\Gamma_\mathrm{MT}$ in the collinear limit, as discussed above. Interestingly, this spin-dependence offers new possibilities for qubit manipulation. Specifically, if by driving $\gMT(\GS)\mathbf{B}$ there is a parallel component to the static spin-quantization axis, the perturbation is longitudinal and only changes the qubit splitting. If it has a transverse component, the spin-quantization axis oscillates and Rabi oscillations between the spin states can be electrically driven. Within the present model, this mechanism therefore does not require a separate Rashba-induced spin-flip tunneling amplitude.

To define the quantization axis at specific measurement points and estimate the driven response, we construct one representative set of $g$-matrices for the superconductor, QD, and magnetotunneling from the measured tensors, as detailed in the Methods and SI section~S13. Within this reconstruction, the magnetotunneling response is strongest for out-of-plane fields, consistent with the heavy-hole character of the states~\cite{katsaros2011spinselective}, but it is not generally collinear with the QD response~\cite{rodriguez2025sweet}.

Using the extracted matrices and how the QD ground-state hybridizes with the superconductor with $\GS$, we calculate the resulting $g$-tensor-modulation ($g$-TMR) Rabi frequency~\cite{CrippaPRL} for a driving field $\delta\GS$. It reaches up to $7.9~\mathrm{MHz}/\mu\mathrm{eV}$ of $\GS$-modulation amplitude at $|\mathbf B|=100~\mathrm{mT}$ for favorable field orientations [\figrefadd{fig:outlook_control_readout}{a-d}]. We note this is a $g$-TMR estimate: static spectroscopy does not determine a possible independent iso-Zeeman contribution associated with a gate-dependent spin-space rotation~\cite{CrippaPRL}.

Magnetotunneling also produces a phase-sensitive response of a resonator in a minimal SC-QD-SC junction extension, enabling spin-readout in a cQED setup. This Josephson junction geometry leads to an effective matrix $\geff(\phiLR)=g_0+g_\varphi\cos\phiLR$ which is characterized by a static $g$-matrix $g_0$ and a magnetotunnel-induced $g$-matrix $g_\varphi$. Crucially, the spin-response now depends on the superconducting phase difference $\phiLR$ even when the spin-orbit strength is negligible. This allows inductive readout by coupling $\phiLR$ to a resonator which converts this phase dependence into a spin-dependent dispersive shift [\figrefadd{fig:outlook_control_readout}{e,f,g}], illustrating how the spin-dependent hybridization enables both electrical spin rotations and dispersive measurements.

The proposed manipulation and readout mechanism, together with the fact that grAl has been successfully implemented in field-resilient cQED experiments \cite{borisov2020superconducting, janik2025strong}, emphasizes the suitability of QDs in planar Germanium as hosts for ASQs with long dephasing times.

\section*{Data Availability}
All data included in this work will be available at the Institute of Science and Technology Austria repository.

\section*{Acknowledgements}
ISTA and ICMM acknowledge support from the European Innovation Council Pathfinder grant no. 101115315 (QuKiT).

The ISTA research was supported by: the Scientific Service Units of ISTA through resources provided by the MIBA Machine Shop, the Nanofabrication, the Electron Microscopy and the Lab Support facility; the NOMIS Foundation and the HE-MSCA-PF project with DOI:10.3030/101150858; the Austrian Science Fund (FWF), research funded in whole or in part [grant DOI, e.g. 10.55776/F86, 10.55776/PAT7682124, 10.55776/P36507]. 
For open access purposes, the author has applied a CC BY public copyright license to any author accepted manuscript version arising from this submission.

The ICMM research is supported by: the Spanish Ministry of Science, Innovation, and Universities through Grants No. CEX2024-001445-S (Severo Ochoa Centres of Excellence program), No. RYC2022-037527-I (Ram\'on y Cajal program), No. PID2022-140552NA-I00, No. PID2023-148257NA-I00, and PID2024-161156NB-I00 funded
by MCIN/AEI/10.13039/501100011033, “ERDF A way
of making Europe” and European Union Next Generation
EU/PRTR; the State Research Agency through the predoctoral Grant No. PRE2022-103741 under the Program
“State Program to Develop, Attract and Retain Talent”; and the Spanish Comunidad de Madrid (CM) “Talento Program” (Project No. 2022-T1/IND-24070).

The ICN2 research is supported by the Severo Ochoa program from Spanish MCIN / AEI (Grant No.: CEX2021-001214-S) and is funded by the CERCA Programme / Generalitat de Catalunya. ICN2 acknowledges funding from Generalitat de Catalunya 2021SGR00457.  This research work has been funded by the European Commission – NextGenerationEU (Regulation EU 2020/2094), through CSIC's Quantum Technologies Platform (QTEP). Authors acknowledge the use of instrumentation as well as the technical advice provided by the Joint Electron Microscopy Center at ALBA (JEMCA). ICN2 acknowledges funding from Grant IU16-014206 (METCAM-FIB) funded by the European Union through the European Regional Development Fund (ERDF), with the support of the Ministry of Research and Universities, Generalitat de Catalunya. ICN2 is founding member of e-DREAM.

ICMM and ICN2 acknowledge support from CSIC Interdisciplinary Thematic Platform (PTI+) on Quantum Technologies (PTI-QTEP+).

\section*{Author contribution}
G.F. and P.F.-S. were responsible for fabricating the devices, conducting measurements and analyzing data under the guidance of G.K.. De.S. and M.B collaborated in the development of the device recipe. M.B., A.B., K.R. and Di.S. developed the oxide layer. K.R. was involved in the development of the grAl recipe and G.F. and A.A.J. in testing it on tunneling spectroscopy and SNS junctions devices. De.S. and A.B. provided assistance during the experiments. I.T., A.G. and J.A. were responsible for the TEM images and EELS analysis, with additional supporting images from T.C.. S.C., D.C. and G.I. grew the Ge QW. D.M.P, R.S.S., M.J.C., R.A., and J.C.A.-U. developed the theory and contributed to data interpretation. D.M.P. and J.C.A.-U. performed the model calculations. G.F., P.F.-S., J.C.A.-U. and G.K. wrote the manuscript with contributions and feedback from all co-authors.

\section*{Competing interests}
The authors declare no competing interests.

\bibliography{references}

\section*{Methods}
\label{sec:methods}

\subsection*{Sample fabrication}
\label{subsec:fab}
 
In this work we focus on two devices, DEV A and DEV B. In both, the Ge/SiGe heterostructure was grown via low-energy plasma-enhanced chemical vapor deposition. The $\text{Si}_{0.3}\text{Ge}_{0.7}$ cap layer thickness varies from 3.5 to 5.5 nm across the wafer; the specific chip used in this work has a 4 nm cap. Growth details are provided in \cite{valentini2024parity}. The charge noise level was found to be  comparable to heterostructures with shallow $ d \sim \SI{20}{\nano \meter}$ capping layers \cite{borovkov2026lownoise}. All structures were defined using electron beam litography (EBL). As the QW is conductive at zero applied gate voltage, we define a mesa region by removing 55 nm of the heterostructure with reactive ion etching using a SF\textsubscript{6}/CHF\textsubscript{3}/O\textsubscript{2} plasma. In DEV A a metallic contact is then added by removing the native SiO\textsubscript{2} cap with an Hf dip and evaporating 20 nm of Pd. Using the same technique we deposit a 20 nm thick grAl film in an electron beam evaporator at an Al rate of around 1 nm/s with a constant oxygen flow of 3.2 sccm for DEV A and 5.5 sccm for DEV B. The evaporation was followed by a static oxidation with a pressure of 5 mbar for 5 mins. Subsequently, Al\textsubscript{2}O\textsubscript{3} was deposited with a thermal atomic layer deposition process at 100$^\circ$C for DEV A and 80$^\circ$C for DEV B. As final step, we defined the Ti/Pd 3/27 nm QD gates.

Alongside DEV A we fabricated grAl stripes 1 and 2 $\mu$m wide. Four probe measurements at 4 K reveal a resistivity of 54 $\mu\Omega$cm and 49 $\mu\Omega$cm respectively. We intentionally chose this relatively low resistivity regime, as Al is known to efficiently proximitize Ge. Through a modest increase in disorder, we aimed to preserve good proximity effect while enhancing magnetic-field resilience. We then increased the resistivity to 170 $\mu\Omega$cm in DEV B to further improve the magnetic-field resilience.

To perform the tunneling spectroscopy of grAl itself (\figrefadd{fig:fig1}{c}), we fabricated a 20 $\mu$m wide stripe at the same resistivity as the grAl used for DEV A and we oxidized the surface in an O\textsubscript{2} atmosphere at a pressure of 5 mbar for 5 mins. Subsequently, we realized a 300 $\mu$m wide Pd tunneling probe overlapping with the grAl stripe.

\subsection*{STEM analysis}
\label{subsec:TEM}
 Xenon base Plasma focused ion beam (PFIB) is used to fabricate electron transparent lamella for the scanning transmission electron microscopy (STEM) analyses, which were conducted using a double-aberration correction Thermo Fisher Scientific (TFS) Spectra 300 microscope operated at 300 keV. Electron energy-loss spectroscopy (EELS) data is obtained by using a Gatan Continuum K3 direct electron detection system with a 0.9 eV/channel dispersion and all the processing were done in Gatan Digital Micrograph software.

\subsection*{Measurement setup}
\label{subsec:setup}
All measurements were performed in a Bluefors LD system with a mixing chamber temperature of 10 mK. Unless otherwise stated, conductance measurements were carried out using a Zurich Instruments MFLI lock-in amplifier at a frequency of 12 Hz. An excitation voltage of 5 $\mu$V was used for zero-field measurements, while a reduced voltage of 2 $\mu$V was applied for measurements in finite magnetic fields to improve the resolution of spin-split states.
The device was measured in a two-probe geometry. The data are presented after subtracting the line resistance of 20 k$\Omega$. When differential conductance is reported via numerical differentiation, the current traces were smoothed using a Savitzky–Golay filter.

\subsection*{Zero-bandwidth model with magnetotunneling}
\label{subsec:zbw}
To model the curves in~\figrefadd{fig:fig3}{} we describe a single QD orbital coupled to a superconductor through a ZBW approximation of the Anderson impurity model, where only one orbital in the superconductor is considered. The model is characterized by the following Hamiltonian:
\begin{equation}
\label{eq:H}
    H_0 = H_{\mathrm{QD}} + H_{\mathrm{S}} + H_{\mathrm{T}} ,
\end{equation}
where:
\begin{equation}
\begin{aligned}
    H_{\mathrm{QD}}&=\frac{U}{2}(n-n_g)^2+H_\text{Z},\\
     H_{\mathrm{S}}&=\xi \sum_\sigma c_\sigma^{\dagger} c_\sigma-\Delta\left(c_{\uparrow}^{\dagger} c_{\downarrow}^{\dagger}+h . c .\right) +H_\text{S,Z},\\
     H_{\mathrm{T}}&=\Gamma_{\mathrm{S}} \sum_\sigma\left(c_\sigma^{\dagger} d_\sigma+h . c .\right)+H_\text{MT}.
\end{aligned}
\end{equation}
$H_{\mathrm{QD}}$ is the Hamiltonian describing the QD, where $d_\sigma$ ($d_\sigma^\dagger$) annihilates (creates) a hole in the QD, $U$ is the charging energy, $n=d_{\uparrow}^{\dagger} d_{\uparrow}+d_{\downarrow}^{\dagger}d_{\downarrow}$, and $n_g$ characterizes the charge offset controlled by the QD's gate voltage $n_g=n_g(V_\mathrm{P})=-\Delta V_\text{P}+2$, where $\Delta V_P$ is the offset we use in~\figrefadd{fig:fig3}{}. Physically, $n_g=1$ sets the boundary between regions I and II, such that $n_g>1$ corresponds to region I and $n_g<1$ corresponds to region II. $H_{\mathrm{S}}$ is the Hamiltonian of the single orbital superconductor, where $c_\sigma$ is the single-particle annihilation operator, $\Delta$ is the superconducting gap and $\xi$ is single-particle energy, which we set to 0 as a reference. $H_{\mathrm{T}}$ is the Hamiltonian describing the coupling between the QD and the superconductor. This coupling is achieved through single-particle spin-preserving tunneling at rate of $\Gamma_{\mathrm{S}}$. 
The Zeeman energies in the quantum-dot region $H_\text{Z}$ and in the superconducting region $H_\text{S,Z}$, are given by
\begin{equation}
    \begin{aligned}
    H_\text{Z}&=\frac{E_\mathrm{Z}}{2}(d_{\uparrow}^{\dagger} d_{\uparrow}-d_{\downarrow}^{\dagger} d_{\downarrow})=\frac{\mu_B}{2}g_{QD}B(d_{\uparrow}^{\dagger} d_{\uparrow}-d_{\downarrow}^{\dagger} d_{\downarrow}) \\
    H_\text{S,Z}&=\frac{E_\mathrm{S,Z}}{2}(c_{\uparrow}^{\dagger} c_{\uparrow}-c_{\downarrow}^{\dagger} c_{\downarrow})=\frac{\mu_B}{2}g_{SC}B(c_{\uparrow}^{\dagger} c_{\uparrow}-c_{\downarrow}^{\dagger} c_{\downarrow}),
    \end{aligned}
\end{equation}
where $g_{SC}$ and $g_{QD}$ are the effective $g$-factors for a given magnetic field orientation in the superconducting and QD regions, respectively. Under the two Zeeman terms, the measured spin splitting is given by a combination of the two effective $g$-factors, depending on the degree of hybridization between the two.

Finally, to account for the asymmetric spin splitting of the doublet state as a function of gate voltage, we consider here the simplest extension to the model that explains the spin splitting together with the observed asymmetries: a spin-dependent tunnel coupling amplitude
\begin{equation}
    H_\text{MT} = \Gamma_\mathrm{MT}(d_\downarrow^\dagger c_\downarrow - d_\uparrow^\dagger c_\uparrow) + \mathrm{H.c.},
    \label{Eq:H_MT}
\end{equation}
where $\Gamma_\mathrm{MT}$ quantifies the asymmetry between the spin-up and spin-down tunneling amplitudes. The distinction between spin up and spin down, implies $\Gamma_\mathrm{MT}$ must break time-reversal symmetry, and, therefore be an odd function of the magnetic field. This magnetotunneling correction has been recently proposed to explain strong $g$-factor renormalizations near charge transitions in hole-based QD arrays~\cite{rodriguez2025sweet}. This term introduces an asymmetry between regions I and II in the spin splitting. Finally, we cast the model into matrix form using a second-quantization package~\cite{secondquantization} and diagonalize it to obtain the curves in~\figrefadd{fig:fig3}{}.

As detailed in the SI section S5, we can approximate the change in Zeeman splitting $\delta E_\text{Z}$ due to the magnetotunneling in the middle of each region~\footnote{We have approximated the middle of each region by $n_g=3/2$ for region I and $n_g=3/2$ for region II.} to be
\begin{equation}
\begin{aligned}
    \delta E_Z^{(\text{I})}&\approx -\frac{2\Gamma_\mathrm{MT}\Gamma_\mathrm{S}}{\sqrt{\Gamma_\mathrm{S}^2+\Delta^2}} \\
    \delta E_Z^{(\text{II})}&\approx +\frac{2\Gamma_\mathrm{MT}\Gamma_\mathrm{S}}{\sqrt{\Gamma_\mathrm{S}^2+\Delta^2}},
    \end{aligned}
\end{equation}
which is opposite for each region.

For the detailed tensorial extension to arbitrary magnetic field orientations of the ZBW model for \figrefadd{fig:outlook_control_readout}{}, see SI.

\clearpage
        
\clearpage
\setcounter{figure}{0}
\renewcommand{\thefigure}{S\arabic{figure}}
\renewcommand{\appendixname}{}
\onecolumngrid
\appendix

\setcounter{section}{0}
\renewcommand{\thesection}{S\arabic{section}}

\counterwithout{equation}{section}   
\renewcommand{\theequation}{S\arabic{equation}}

\setcounter{equation}{0}

\begin{center}
{\Large\bfseries Supplementary Information\par}
\end{center}

\section{Extended Transmission Electron Microscopy images}
\label{sec:enG}

\begin{figure*}[!h]
    \centering
    \includegraphics[width = \linewidth]{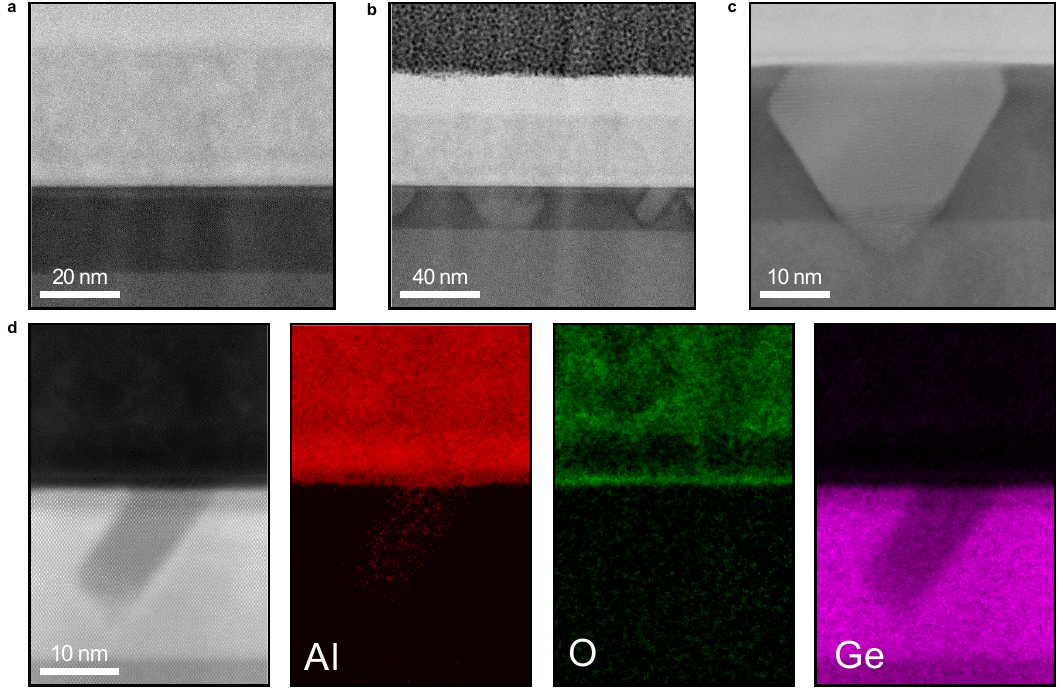}
        \caption{\textbf{STEM and EELS analysis of Al-rich defects in a Ge/SiGe heterostructure.} \textbf{a-c.} Bright-field scanning transmission electron microscopy (BF-STEM) images of a Ge/SiGe heterostructure with a grAl layer on top of a Ge quantum well (QW), showing a defect-free region (a) and regions containing defects in the QW (b,c). The defects exhibit sharp interfaces with the QW and are bounded by facets inclined at roughly 60° with respect to the QW plane. Notably, this angle coincides with the typical dislocations propagation direction in SiGe systems. \textbf{d.} High-angle annular dark-field STEM and the corresponding electron energy-loss spectroscopy (EELS) analysis of a region with a defect in the QW, revealing that the defect is an aluminium-rich crystal embedded inside the germanium and silicon-germanium layers. We speculate that aluminium diffuses from the grAl layer into the QW along structural defects. Possibly, this diffusion happens during the atomic layer deposition of the oxide layer.}
    
    \label{fig:figS0}
\end{figure*}

\newpage 

\section{Extended tunneling spectroscopy data}
\label{sec:extTS}
\begin{figure*}[!h]
    \centering
    \includegraphics[width = \linewidth]{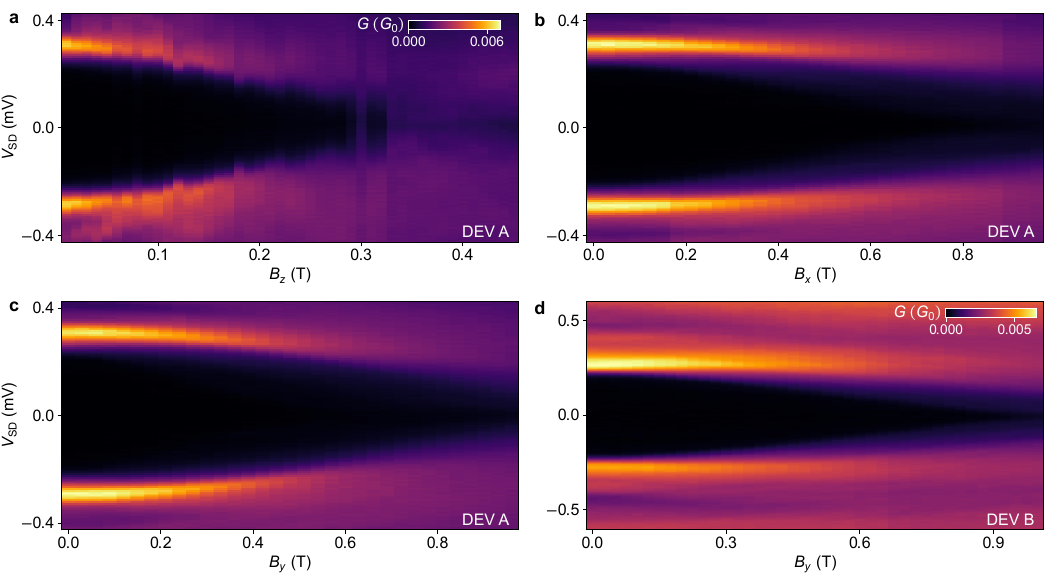}
    \caption{\textbf{Field resilience of DEV A.} \textbf{a-c.} Differential conductance $G$ = $dI/dV$ of DEV A in units of $\mathrm{G_0 = 2e^2/h}$ vs $V_{\mathrm{SD}}$ and magnetic field $B_{z,x,y}$ at $V_\mathrm{B}$ = 4.35V and $V_\mathrm{S}$ = 3.30V. A region of reduced subgap conductance is visible up to 300 mT in the out-of-plane ($z$) direction and above 950 mT in-plane. \textbf{d.} $G$ of DEV B vs $V_{\mathrm{SD}}$ and $B_y$ in the same gate configuration as in Figure 1.f,g. }
    
    \label{fig:figS11}
\end{figure*}

\begin{figure*}[!h]
    \centering
    \includegraphics[width = \linewidth]{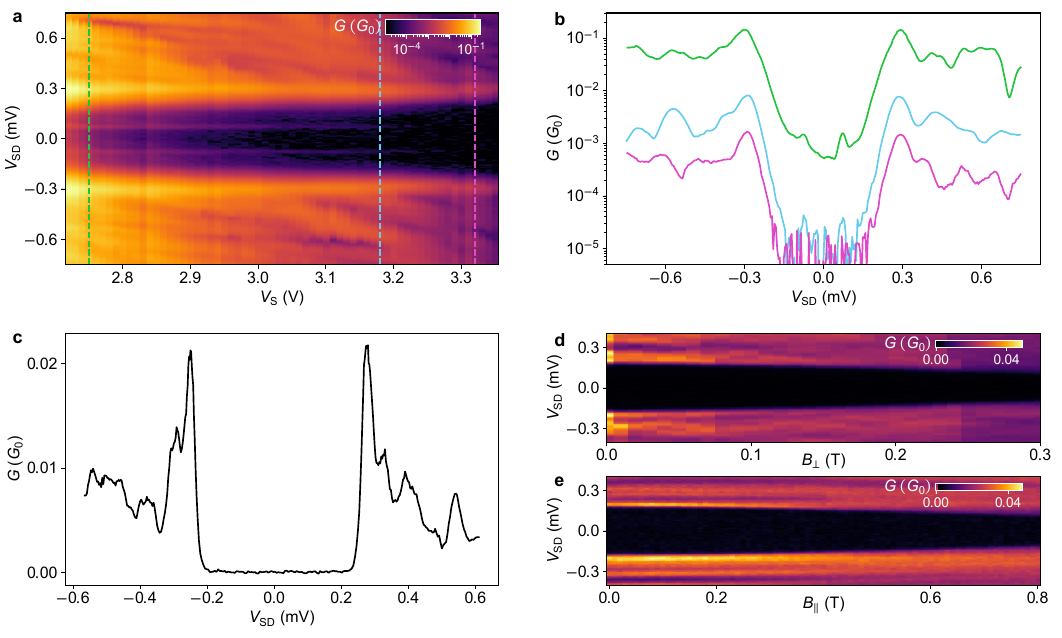}
    \caption{\textbf{Extended tunneling spectroscopy data.} \textbf{a.}  Differential conductance $G$ = $dI/dV$ in units of $\mathrm{G_0 = 2e^2/h}$ and in logarithmic scale as a function of the voltage applied to gate S ($V_{\mathrm{S}}$) and bias voltage ($V_{\mathrm{SD}}$), at $V_\mathrm{B}$ = 4.6V, 250 mV higher than in Fig. 1b. \textbf{b.} Line-cuts from (a)
    at gate voltage indicated by the dashed lines. 
    \textbf{c.} $G$ vs $V_\mathrm{SD}$ for a second equivalent device on the same chip, where we observe BCS peaks at 263 $\mu$V. Here, we obtained $G$ by numerically differentiating the measured current after a smoothing procedure (Savitzky–Golay filter) is applied and a line resistance of 923 k$\Omega$ is subtracted. \textbf{d-e.} G as a function of $V_\mathrm{SD}$ and out-of and in-plane magnetic fields $B_{\perp,\parallel}$, for a third tunneling spectroscopy device. The gapped region exceeds 300 mT of resilience in the perpendicular direction and 800 mT in the parallel one. Here, BCS peaks are at $\pm$200 $\mu$V. Here, the superconduting contact has a width of 3 $\mu$m and a thickness of 23 nm.}
    
    \label{fig:figS1}
\end{figure*}

In this supplementary section, we present additional tunneling spectroscopy measurements that illustrate the variability in the induced superconducting gap and its magnetic field resilience. In \figrefadd{fig:figS11}{}, we report how the induced supercondcuting gap of DEV A evolves in magnetic field in the three magnet directions. We measure conductance below $\mathrm{6}\times\mathrm{10^{-5}}$ $G_0$ (noise level) at zero bias up to 160 mT in the out-of-plane direction and 800 mT in-plane for a 500 nm wide and 20 nm thick grAl contact. In \figrefadd{fig:figS1}{a}, we show tunneling spectroscopy as a function of $V_\mathrm{S}$ taken at a $V_\mathrm{B}$ value different from that used in the main text. The corresponding line cuts at the gate voltages marked by the colored dashed lines are shown in \figrefadd{fig:figS1}{b}.
In \figrefadd{fig:figS1}{c}, we show an equivalent measurement for a second device on the same chip, revealing BCS peaks at $|V_{SD}| = 263~\mu\text{V}$. For earlier devices, such as those in \figrefadd{fig:figS1}{d–e}, we measured superconducting gaps of approximately 200 $\mu$eV. We note that these earlier devices were fabricated using a different grAl evaporation procedure, characterized by longer pumping times to reach the evaporation pressure and by a substantially longer O\textsubscript{2} exposure (roughly three times longer) prior to initiating the evaporation. In earlier devices we have also varied the dielectric layer, changing the type of oxide (Al\textsubscript{2}O\textsubscript{3} or HfO\textsubscript{2}), the deposition temperature and the deposition duration. Across these devices we measured variations in the magnetic field resilience, which might be associated with the local diffusion of Al in the Ge QW, as shown in \figrefadd{fig:figS0}{}.
\newpage
\newpage

\newpage

\section{Enhanced conductance}
\label{sec:enG}

\begin{figure*}[!h]
    \centering
    \includegraphics[width = \linewidth]{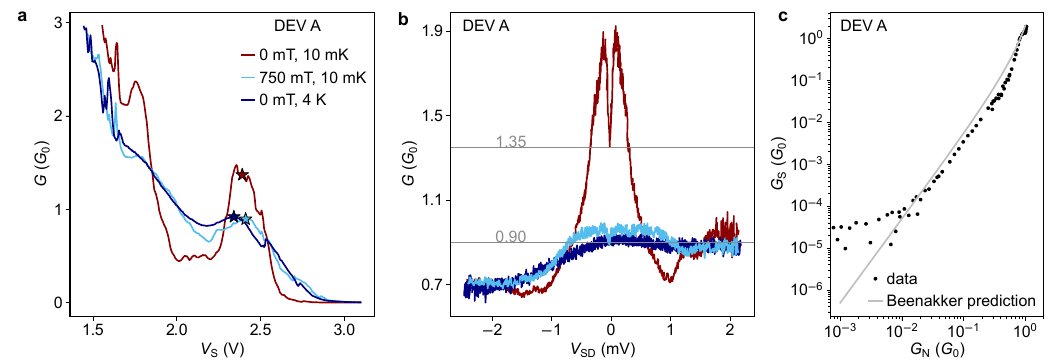}
    \caption{\textbf{Andreev induced enhanced sub gap conductance of DEV A.} \textbf{a.}  Zero bias differential conductance $G$ = $dI/dV$ in units of $\mathrm{G_0 = 2e^2/h}$ as a function of gate voltage $V_\mathrm{S}$ at $V_\mathrm{B}$ = 4.35 V, as in Fig. 1 of main text. The red trace is taken at 0 field and base temperature, whereas the light-blue and the blue traces are taken with the superconductor turned normal by applying 750 mT out-of-plane (light-blue) or going to 4 K (blue). At high $V_\mathrm{S}$ we observe suppressed zero bias conductance in the superconducting state, whereas around $V_\mathrm{S}$ = 2.4 V, where the conductance in the normal state approaches $\mathrm{G_0}$, we observe enhanced conductance. We note that the blue trace is slightly shifted  to the right compared to the light-blue trace likely due to some small hysteresis.  \textbf{b.} Differential conductance $G$ vs bias $V_{\mathrm{SD}}$ for gate voltages $V_\mathrm{S}$ indicated by the star markers in (a). Around zero bias, the red trace (superconducting state) shows enhanced conductance approaching twice the conductance quantum, as expected from Andreev reflection. For a single, perfectly transparent channel, $G$ reaches 2$G_0$. A peak reaching 2$G_0$ is expected for a single, perfectly transparent channel, while reduced transparency leads to a dip near zero bias \cite{Cuevas1996}. The observed dip reaches $G$ = 1.35 $G_0$. Assuming a single spin degenerate channel, the BTK model gives the Andreev-enhanced conductance as $G_\mathrm{S} = \frac{4e^2}{h}\frac{T^2}{(2-T)^2}$, which yields $1.35 = 2\frac{T^2}{(2-T)^2}$. Solving for $T$ gives $T = 0.9$, matching the zero bias normal state conductance. \textbf{c.} Zero-bias conductance in the superconducting state ($G_S$) plotted against the corresponding normal-state value ($G_N$). The normal-state conductance ($G_N$) is obtained by locally heating the chip above the grAl critical temperature using an on-chip resistor. After the device cools back down, the superconducting conductance ($G_S$) is measured under identical gate conditions. Each data point corresponds to a different $V_\mathrm{S}$ value, while $V_\mathrm{B}$ is fixed. The data points show a slight discrepancy from the theoretical prediction for a single channel $G_\mathrm{S} = 2G_0\frac{G_\mathrm{N}^2}{(2G_0-G_\mathrm{N})^2}$. A possible explanation for the deviation is the formation of multiple channels. The measurements in panel C were taken during a different cooldown than the other measurements of DEV A.} 
    \label{fig:figS2}
\end{figure*}

\newpage 

\section{Replicas}
\label{sec:replicas}
\begin{figure*}[!h]
    \centering
    \includegraphics[width = \linewidth]{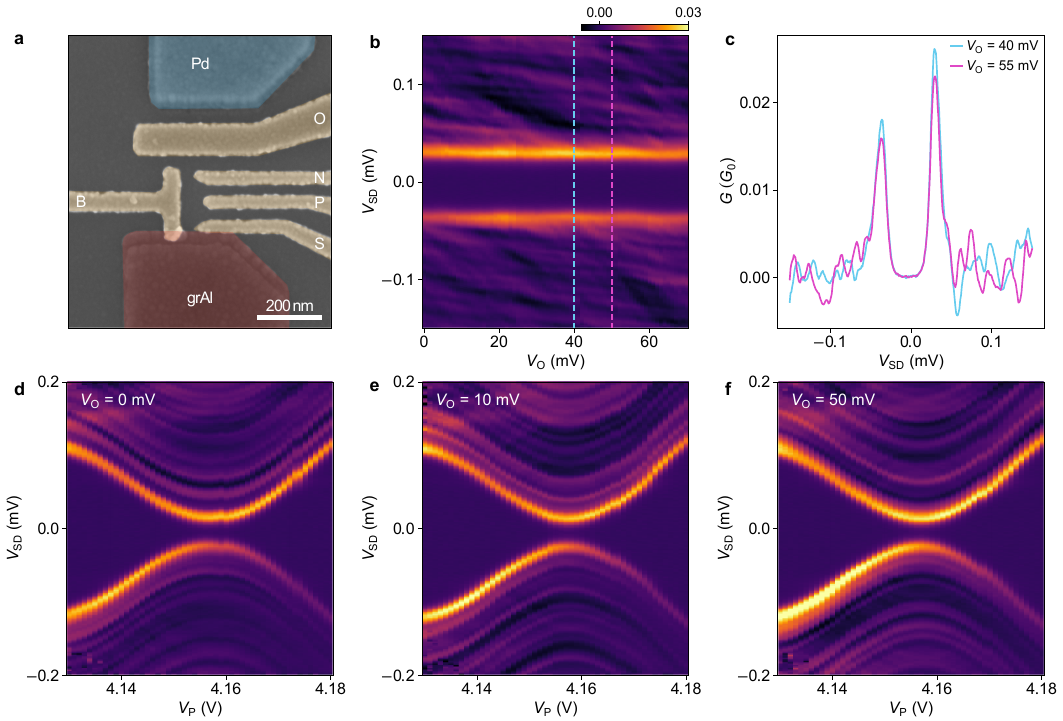}
    \caption{\textbf{Replicas of sub-gap states} \textbf{a.} False-colour scanning electron microscopy image of a copy of the device, with the superconducting grAl contact in red, the Pd contact in blue and the gates to form the quantum dot in yellow. Gate O has the purpose of tuning the local density of states close to the dot. \textbf{b.} $G$ vs $V_\mathrm{O}$ and $V_\mathrm{SD}$ measured with a lock-in amplifier at a fixed plunger voltage $V_\mathrm{P}$ = 4151 mV. The sub-gap states stay at their position whereas at higher bias the conductance map changes smoothly with $V_\mathrm{O}$. \textbf{c.} Linecuts at $V_\mathrm{O}$ =  40 mV and $V_\mathrm{O}$ = 50 mV indicated by the light-blue and purple dashed lines in panel b, respectively. \textbf{d-f}. $G$ vs. $V_\mathrm{P}$ and $V_\mathrm{SD}$ for $V_\mathrm{0} $ = 0 mV, 10 mV, 50 mV. The carpet of peaks follows the high conductance sub-gap states throughout the entire 2D map smoothly. The position and conductance of the replicas is changed with $V_\mathrm{0}$.}
    \label{fig:figS3}
\end{figure*}

For all measurements, we observed replicas of the sub-gap states and negative differential conductance, as shown in \figrefadd{fig:figS3}{b-f}. Similar results in tunneling spectroscopy measurements were previously attributed to a non-uniform local density of states (LDOS) of the Ge lead \cite{schmidt1997observation, schmidt2001energy}. To investigate the origin of the effect in our measurements, we electrostatically tuned the LDOS of the lead with the gate O (\figrefadd{fig:figS3}{a}) and observed a clear change of the replica pattern while the sub-gap states stayed unaffected, as shown in \figrefadd{fig:figS3}{b}. 

\newpage

\section{Asymmetric spin splitting}
\label{sec:magnetotunnel}
In Fig. 3 from the main text, we observe an asymmetric spin splitting of the doublet state as a function of gate voltage. Here, we evaluate in detail how the zero-bandwidth (ZBW) model we describe in the Methods section reproduces the asymmetric spin splitting after including the asymmetric tunneling. To make this section self-contained, first we remind our ZBW model:
\begin{equation}
\label{eq:H}
    H_0 = H_{\mathrm{QD}} + H_{\mathrm{S}} + H_{\mathrm{T}} ,
\end{equation}
where:
\begin{equation}
\begin{aligned}
    H_{\mathrm{QD}}&=\frac{U}{2}(n-n_g)^2+\frac{\mu_B}{2}g_{QD}B(d_{\uparrow}^{\dagger} d_{\uparrow}-d_{\downarrow}^{\dagger} d_{\downarrow}),\\
     H_{\mathrm{S}}&=\xi \sum_\sigma c_\sigma^{\dagger} c_\sigma-\Delta\left(c_{\uparrow}^{\dagger} c_{\downarrow}^{\dagger}+h . c .\right) +\frac{\mu_B}{2}g_{SC}B(c_{\uparrow}^{\dagger} c_{\uparrow}-c_{\downarrow}^{\dagger} c_{\downarrow}),\\
     H_{\mathrm{T}}&=\Gamma_{\mathrm{S}} \sum_\sigma\left(c_\sigma^{\dagger} d_\sigma+h . c .\right)+H_\text{MT},
\end{aligned}
\end{equation}
where $d_\sigma$ ($d_\sigma^\dagger$) annihilates (creates) a hole in the QD, and
$
n=d_{\uparrow}^{\dagger}d_{\uparrow}+d_{\downarrow}^{\dagger}d_{\downarrow},
$
with charging energy $U$ and gate-controlled offset charge $n_g=n_g(V_P)$.
The superconducting term $H_{\mathrm S}$ is a single-orbital model with annihilation operator $c_\sigma$, gap $\Delta$, and single-particle energy $\xi$, set to $\xi=0$ as reference.
The coupling term describes spin-conserving QD-superconductor (SC) tunneling with amplitude $\Gamma_{\mathrm S}$.
Zeeman terms are parametrized by effective $g$-factors $g_{\mathrm{QD}}$ and $g_{\mathrm{SC}}$ (for the chosen magnetic-field orientation) in the QD and SC regions, respectively.
As the observations do not require non-collinear spin quantization axes between QD and SC nor a spin-flip tunneling  mechanism, we neglect such terms.

Finally, we introduce an asymmetry in the spin splitting as a spin-dependent tunnel coupling amplitude that was recently proposed for quantum dot arrays with hole spins~\cite{rodriguez2025sweet}:
\begin{equation}
    H_\text{MT} = \Gamma_\mathrm{MT}(d_\downarrow^\dagger c_\downarrow - d_\uparrow^\dagger c_\uparrow) + \mathrm{H.c.},
    \label{Eq:H_MT}
\end{equation}
where $\Gamma_\mathrm{MT}$ quantifies the asymmetry between the spin-up and spin-down tunneling amplitudes. $\Gamma_\mathrm{MT}$ must break time-reversal symmetry, and, therefore be an odd function of the magnetic field. Interestingly, adding $H_\text{MT}$ as a correction to the ZBW model introduces an asymmetry in the spin splitting. 

To understand how the asymmetry emerges, we analyze the spin-doublet sector. Since the spin splitting is defined within the doublets, and the singlet states are decoupled from them, we can restrict the discussion to the odd subspaces. Moreover, neglecting spin-flip tunneling and non-collinear spin terms makes the Hamiltonian block-diagonal, with the spin-projection in the magnetic field direction being a good quantum number. In the minimal basis
\[
\mathcal{B}_{+}=\{\ket{\uparrow\downarrow_{QD},\uparrow_{SC}},\ket{\uparrow_{QD},\uparrow\downarrow_{SC}},\ket{\uparrow_{QD},0_{SC}},\ket{0_{QD},\uparrow_{SC}}\},
\]
which spans the spin-doublet \(S_z=+1/2\) manifold, the corresponding \(S_z=-1/2\) manifold is generated by time reversal, \(\mathcal{B}_{-}=\mathcal{T}\mathcal{B}_{+}\). The block Hamiltonians (up to identity terms) are

\begin{equation}
    H_{\pm 1/2}=\begin{pmatrix}
        2U(1-n_g) & \Gamma_S\pm \Gamma_\text{MT} & 0 & 0 \\
        \Gamma_S\pm \Gamma_\text{MT} & U(1/2-n_g) & -\Delta & 0 \\
        0 & -\Delta & U(1/2-n_g) & -\Gamma_S\pm \Gamma_\text{MT} \\
        0 & 0 & -\Gamma_S\pm \Gamma_\text{MT} & 0
    \end{pmatrix} \pm \frac{\mu_B B}{2} \text{diag}(g_\text{SC},g_\text{QD},g_\text{QD},g_\text{SC}).
    \label{eq:Hodd}
\end{equation}
Where, $n_g$ is the dimensionless gate parameter that parametrizes the gate voltage. The center of the doublet-ground-state region is at $n_g=1$; region I corresponds to $n_g>1$, while region II corresponds to $n_g<1$ (see Fig. 3 of main text). From here, we can estimate the Zeeman splitting of the doublet ground state. Exact diagonalization in the general case is not possible. However, there are three values of $n_g=(1/2,1,3/2)$ with higher symmetry that enable partial diagonalization of the doublet Hamiltonians. In this way, we estimate the Zeeman splitting as
\begin{equation} \label{eq:zsplitting}
\begin{aligned}
    & n_g \approx 1 : \qquad &&E_\text{Z}^\text{eff} = \frac{\mu_BB}{2} \left(g_\mathrm{QD} + g_\mathrm{SC} + \frac{(g_\mathrm{QD} - g_\mathrm{SC})(2\Delta + U)}{\sqrt{
   16 \Gamma_S^2 + (2\Delta + U)^2}} \right)-\frac{16\Gamma_S\Gamma_\text{MT}U}{(U+2\Delta)^2}(n_g-1)
   \\
   & n_g \approx 1/2 : \qquad &&E_\text{Z}^\text{eff} \approx \mu_BB\left(g_\mathrm{QD}\sin^2\theta + \frac{1}{2}(g_\mathrm{QD} + g_\mathrm{SC})\cos^2\theta\right) + 2\Gamma_\mathrm{MT}\cos\theta
   \\
   & n_g \approx 3/2 : \qquad &&E_\text{Z}^\text{eff} \approx \mu_BB\left(g_\mathrm{QD}\sin^2\theta + \frac{1}{2}(g_\mathrm{QD} + g_\mathrm{SC})\cos^2\theta\right) - 2\Gamma_\mathrm{MT}\cos\theta
   \\
   & && \theta = \arctan\frac{\Delta}{\Gamma_S}.
\end{aligned}
\end{equation}
These equations clearly show that the asymmetry in spin splitting about $n_g=1$ is directly proportional to the magnetotunneling term $\Gamma_\text{MT}$. For the illustrative cases $n_g=\{1/2,3/2\}$ of regions II and I respectively, which would have the same effective Zeeman splitting without magnetotunnel, the difference is given by $4\Gamma_\text{MT}\cos\theta$.

\subsection{Physical origin of the magnetotunneling term}
In the Ge hole device considered here (and in particular in the many-body regime), the relevant bound states are generically not pure heavy-hole (HH) or light-hole (LH) states, but HH-LH hybrids. This hybridization is primarily kinetic in origin and is naturally captured by the Kohn–Luttinger Hamiltonian. In addition, since valence-band holes carry spin $3/2$, magnetic interactions do not simply act on a fixed two-level “spin” subspace: they can also modify, and be modified by, the HH-LH admixture. A key manifestation is the strong anisotropy contrast between HH and LH effective $g$-factors: HHs typically feature a large vertical $g$-factor and much smaller in-plane $g$-factors, whereas LHs are less anisotropic and can exhibit comparatively stronger in-plane $g$-factors than vertical ones. Since the HH-LH mixing is momentum dependent, the effective $g$-factors of an arbitrary bound state inherit an intrinsic momentum dependence~\cite{winkler2003spin}. At the level of an effective two-level description for a specific bound state, this can be summarized as a Zeeman term of the form
\begin{equation}
    H^{\text{eff}}_Z \propto \boldsymbol{\sigma}\cdot g(p)\mathbf{B},
\end{equation}
where $g(p)$ denotes the (generally anisotropic) effective $g$-tensor evaluated for the momentum content of the bound state. Note that symmetry arguments require $g$ to be an even function of $p$.

Tunneling between localized sites is itself a kinetic effect. In a minimal two-site toy model with localized states $\ket{L}$ and $\ket{R}$, the dominant contribution to the tunnel coupling is set by the kinetic-energy operator,
\begin{equation}
    H^{\text{eff}}_t \approx \bra{L}\frac{p^2}{2m}\ket{R}.
\end{equation}
This term is precisely what yields a non-negligible coupling between spatially separated sites. It is therefore natural to expect that a momentum-dependent Zeeman interaction produces an analogous off-diagonal matrix element in the presence of a magnetic field, namely
\begin{equation}
    H^{\text{eff}}_\text{MT} \approx \bra{L}\boldsymbol{\sigma}\cdot g(p)\mathbf{B}\ket{R},
\end{equation}
which constitutes the magnetotunneling correction to the effective Hamiltonian. Because this contribution is controlled by momentum dependence (in practice dominated by $p^2$), it is expected to scale with the same kinetic overlaps that generate the bare tunnel coupling, and is thus typically proportional to the tunnel amplitude, i.e., $\Gamma_\text{MT}\propto \Gamma_S$. Moreover, since the momentum dependence of $g(p)$ is strongly tied to HH-LH mixing, the effect should be highly sensitive to the HH-LH energy separation, becoming particularly pronounced when HH and LH configurations are brought close in energy. 
Spin-selective tunneling of this origin was reported in Ge-based hole QDs~\cite{katsaros2011spinselective}, where it was strongly anisotropic--pronounced for out-of-plane fields and nearly absent in-plane--and where the spin selectivity was already characterized by a direction in spin space rather than by a scalar. The formulation of Ref.~\cite{rodriguez2025sweet} promotes this to a full tunnel $g$-matrix, which is the object we extract in \secref{sec:sm_tensor_extraction}.

Finally, while HH-LH mixing is the relevant and physically transparent mechanism in our specific device, it should not be viewed as the only route to magnetotunneling physics. More broadly, any high-spin-orbit system that exhibits momentum-dependent effective $g$-factors is susceptible to analogous corrections, including electron platforms with strong spin-orbit coupling~\cite{electron_g-factor}.

\subsection{Connection with continuum superconducting models}\label{subsec:continuum}
The previous description is based on a ZBW model, where the superconductor is approximated as an orbital level and, therefore, the parameters of such model must only be considered as qualitative rather than quantitative. To connect the ZBW description with a realistic superconducting continuum, we integrate out the BCS lead and compare the resulting dot self-energy with its ZBW counterpart. This procedure clarifies why the effective tunneling parameters of the ZBW model are renormalized quantities.

We write the full continuum Hamiltonian in dot/lead blocks as
\begin{equation}
\tilde{H}=
\begin{pmatrix}
\tilde{H}_{\mathrm{QD}} & \tilde{H}_{\mathrm{QD-SC}}\\
\tilde{H}_{\mathrm{SC-QD}} & \tilde{H}_{\mathrm{SC}}
\end{pmatrix},
\end{equation}
where $\tilde{H}_{\mathrm{QD}}$ is the isolated dot Hamiltonian, $\tilde{H}_{\mathrm{SC}}$ is the BCS lead Hamiltonian, and $\tilde{H}_{\mathrm{QD-SC}}=\tilde{H}_{\mathrm{SC-QD}}^\dagger$ is the coupling block.

To make the origin of the couplings explicit, we parametrize
\begin{equation}
\tilde{H}_{\mathrm{QD-SC}}=\sum_k \Psi_d^\dagger \tilde{V}_k \Psi_k+\mathrm{H.c.},
\qquad
\tilde{V}_k=\tau_z\!\left(\tilde{\Gamma}_{S,k}\sigma_0-\tilde{\Gamma}_{\mathrm{MT},k}\sigma_3\right),
\label{eq:Vk_def}
\end{equation}
with $\Psi_d=(d_\uparrow, d_\downarrow, d_\uparrow^\dagger, d_\downarrow^\dagger)$ and $\Psi_k=(c_{\uparrow,k}, c_{\downarrow,k}, c_{\uparrow,k}^\dagger, c_{\downarrow,k}^\dagger)$ being the dot and lead Nambu spinors. In this basis,
\begin{equation}
\tilde{V}_k=
\begin{pmatrix}
\tilde{\Gamma}_{S,k}-\tilde{\Gamma}_{\mathrm{MT},k} & 0 & 0 & 0\\
0 & \tilde{\Gamma}_{S,k}+\tilde{\Gamma}_{\mathrm{MT},k} & 0 & 0\\
0 & 0 & -\tilde{\Gamma}_{S,k}+\tilde{\Gamma}_{\mathrm{MT},k} & 0\\
0 & 0 & 0 & -\tilde{\Gamma}_{S,k}-\tilde{\Gamma}_{\mathrm{MT},k}
\end{pmatrix}.
\end{equation}

Using the retarded Green function ($\omega\to\omega+i0^+$), the dot block satisfies
\begin{equation}
G_{\mathrm{QD}}^{-1}(\omega)=\omega-\tilde{H}_{\mathrm{QD}}-\Sigma(\omega),\qquad
\Sigma(\omega)=\tilde{H}_{\mathrm{QD-SC}}(\omega-\tilde{H}_{\mathrm{SC}})^{-1}\tilde{H}_{\mathrm{SC-QD}}.
\end{equation}
Equivalently, mode by mode,
\begin{equation}
\Sigma(\omega)=\sum_k \tilde{V}_k\,(\omega-\tilde{H}_{\mathrm{SC},k})^{-1}\,\tilde{V}_k^\dagger.
\end{equation}

In Nambu$\otimes$spin space (Pauli matrices $\tau_i$ in Nambu space and $\sigma_i$ in spin space), for
$\tilde{H}_{\mathrm{SC},k}=\epsilon_k\tau_z\sigma_0+\Delta\tau_y\sigma_y$, one obtains
\begin{equation}
\Sigma(\omega)=\sum_k
\frac{\omega\,\tilde{T}_\omega^{(k)}+\epsilon_k\,\tilde{T}_\epsilon^{(k)}+\Delta\,\tilde{T}_\Delta^{(k)}}
{\omega^2-\epsilon_k^2-\Delta^2},
\end{equation}
where $\epsilon_k$ is the normal-state dispersion and $\Delta$ is the superconducting gap. The relevant structures are
\begin{align}
\tilde{T}_\omega^{(k)}&=(\tilde{\Gamma}_{S,k}^2+\tilde{\Gamma}_{\mathrm{MT},k}^2)\tau_0\sigma_0
-2\tilde{\Gamma}_{S,k}\tilde{\Gamma}_{\mathrm{MT},k}\tau_0\sigma_3,\\
\tilde{T}_\Delta^{(k)}&=(\tilde{\Gamma}_{S,k}^2-\tilde{\Gamma}_{\mathrm{MT},k}^2)\tau_y\sigma_y.
\end{align}
For a symmetric band, the $\epsilon_k$ term is odd and vanishes after energy integration.

Assuming a flat normal-state density of states $\rho$ around the Fermi level and $k$-independent couplings,
$\tilde{\Gamma}_{S,k}\equiv\tilde{\Gamma}_S$ and
$\tilde{\Gamma}_{\mathrm{MT},k}\equiv\tilde{\Gamma}_{\mathrm{MT}}$, we get
\begin{equation}
\Sigma_{\mathrm{cont}}(\omega)=
-\frac{\pi\rho}{\sqrt{\Delta^2-\omega^2}}
\left[
\omega\!\left(\tilde{\Gamma}_S^2+\tilde{\Gamma}_{\mathrm{MT}}^2\right)\tau_0\sigma_0
-2\omega\,\tilde{\Gamma}_S\tilde{\Gamma}_{\mathrm{MT}}\tau_0\sigma_3
+\Delta\!\left(\tilde{\Gamma}_S^2-\tilde{\Gamma}_{\mathrm{MT}}^2\right)\tau_y\sigma_y
\right].
\label{eq:sigma_cont_final}
\end{equation}

In the ZBW model, the continuum is replaced by one effective superconducting level at energy $\epsilon_0$, with effective couplings $\Gamma_S$ and $\Gamma_{\mathrm{MT}}$:
\begin{equation}
\Sigma_{\mathrm{ZBW}}(\omega)=
\frac{
\omega\!\left(\Gamma_S^2+\Gamma_{\mathrm{MT}}^2\right)\tau_0\sigma_0
-2\omega\,\Gamma_S\Gamma_{\mathrm{MT}}\tau_0\sigma_3
+\Delta\!\left(\Gamma_S^2-\Gamma_{\mathrm{MT}}^2\right)\tau_y\sigma_y
}{
\omega^2-\epsilon_0^2-\Delta^2
}.
\label{eq:sigma_zbw_final}
\end{equation}

Matching Eqs.~(\ref{eq:sigma_cont_final}, \ref{eq:sigma_zbw_final}) at low energy ($\omega=0$) and for $\epsilon_0=0$ gives
\begin{equation}
\Gamma_S^2-\Gamma_{\mathrm{MT}}^2
=\pi\rho\Delta\left(\tilde{\Gamma}_S^2-\tilde{\Gamma}_{\mathrm{MT}}^2\right).
\end{equation}
If both channels renormalize with the same prefactor, then
\begin{equation}
\Gamma_S\simeq \sqrt{\pi\rho\Delta}\,\tilde{\Gamma}_S,
\qquad
\Gamma_{\mathrm{MT}}\simeq \sqrt{\pi\rho\Delta}\,\tilde{\Gamma}_{\mathrm{MT}}.
\end{equation}

Hence, $\Gamma_S$ and $\Gamma_{\mathrm{MT}}$ in ZBW, Eqs.~(\ref{Eq:H_MT},\ref{eq:H}), fits should be interpreted as effective DOS-renormalized parameters rather than bare microscopic tunnel amplitudes. An intrinsic spin-dependence of the DOS may further enhance the magnetotunnel effect.

\newpage

\section{Fitting procedure for YSR}
\begin{figure*}[h!]
    \centering
    \includegraphics[width = \linewidth]{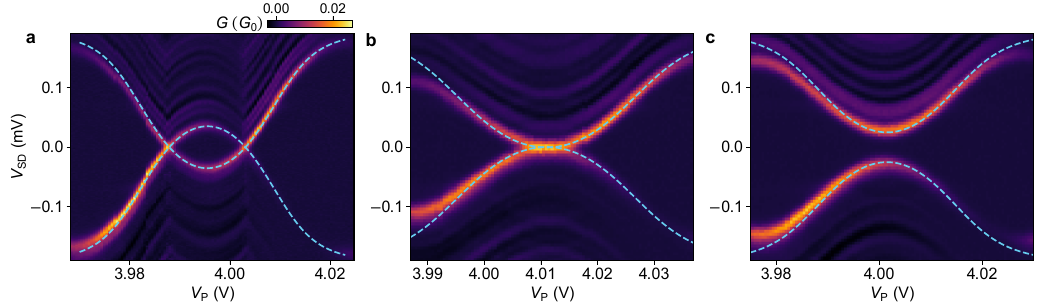}
    \caption{\textbf{ZBW model fit of YSR state.} \textbf{a-c.} $G$ vs $V_\mathrm{P}$ and $V_{\mathrm{SD}}$ with increasing $\Gamma_{\mathrm{S}}$. Same measurements as those shown in Fig. 2c-e. The dashed lines correspond to fits obtained through a ZBW model, from which we extract a charging energy of 685 $\mu$eV and couplings to the superconductor $\Gamma_\mathrm{S}$ = 156 $\mu$eV (c), 183 $\mu$eV (d), 204 $\mu$eV.}
    
    \label{fig:figSzbwfit}
\end{figure*}

Conductance peaks observed in transport correspond to transitions from the ground state to excited states of opposite parity. Accordingly, we fit the features shown in Fig. 2c–e to the difference between the two lowest-energy eigenvalues of Hamiltonian \ref{eq:H} with opposite parities. For all three coupling configurations, we fix $\Delta = 200~\mu\text{eV}$, consistent with the strongly suppressed conductance region observed in tunneling spectroscopy. For the lowest coupling configuration (Fig. 2c), we use $\Gamma_{\mathrm{S}}$, $U$, and the lever arm as free parameters, from which we extract values of 156 $\mu$eV, 685 $\mu$eV, and 0.028, respectively. We do not use the parameters extracted from the Coulomb diamonds in Fig. 2a, as the gate configuration has been substantially modified. In particular, gate barrier S has been significantly lowered (from 3.98 V to 2.60 V), reducing both the charging energy and the plunger-gate lever arm. For the highest coupling configuration (Fig. 2e), we fix $U$ and the lever arm to the values obtained for $\Gamma_{\mathrm{S}} = 156~\mu\text{eV}$, leaving $\Gamma_{\mathrm{S}}$ as the only free parameter. This yields $\Gamma_{\mathrm{S}} = 204~\mu\text{eV}$. To obtain a good fit in the intermediate coupling regime (Fig. 2d), we again treat the lever arm and $\Gamma_{\mathrm{S}}$ as free parameters while fixing $U = 685~\mu\text{eV}$. From this, we extract $\Gamma_{\mathrm{S}} = 183~\mu\text{eV}$ and a lever arm of 0.025, close to the value obtained for the lowest coupling. We note that stabilizing the device in this intermediate regime required adjusting not only gate S but also the backbone gate B by 20 mV.\\
Finally, we note that all fits were performed for $|V_{\mathrm{SD}}| < 100~\mu\text{V}$. In a single QD orbital ZBW model, the transition energy approaches the superconducting gap. In the multi-orbital regime, the transition energy can be smaller than the superconducting gap, depending on the relative magnitudes of $\Delta$, $\Gamma_{\mathrm{S}}$, and the dot addition energy. This results in deviations from the single QD orbital ZBW-model prediction, as is evident in \figrefadd{fig:figSzbwfit}{}.

\newpage

\subsection{Model-experiment comparison at finite magnetic fields}

\begin{figure*}[h!]
    \centering
    \includegraphics[width=\linewidth]{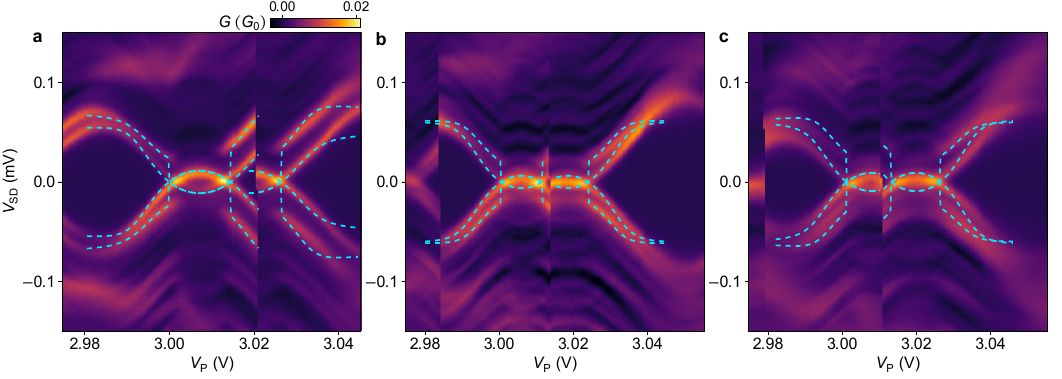}
    \caption{\textbf{Qualitative model-experiment comparison for a YSR state at finite magnetic field.} \textbf{a-c.} $G$ vs $V_\mathrm{P}$ and $V_\mathrm{SD}$ at $B_z$= 100 mT (a), $B_x$= 350 mT (b) and $B_y$= 350 mT (c). Dashed lines correspond to theoretical results obtained through the ZBW model in \eqref{eq:H} using $\Delta$ = 61 $\mu$eV, $U$ = 610 $\mu$eV, $\Gamma_\mathrm{S}$ = 90 $\mu$eV and incorporating magnetotunnel couplings $\Gamma_\mathrm{MT}=0.09\Gamma_\mathrm{S}$ (a), $\Gamma_\mathrm{MT}=0.00\Gamma_\mathrm{S}$ (b), $\Gamma_\mathrm{MT}=-0.04\Gamma_\mathrm{S}$ (c). The $g$-factors used in the model are $g_\mathrm{QD}$ = 5 (a), 1.1 (b), 1.5 (c) and $g_\mathrm{SC}$ = 3.5 (a), 0.0 (b), 0.0 (c). Note these are effective fitting parameters for this specific dataset and must be considered qualitative rather than the real microscopic parameters. We observe charge switches in each plot. These measurements were acquired during a different cooldown than the other measurements of DEV A.
    }
    \label{fig:figSfit}
\end{figure*}

By applying a magnetic field, the YSR states are spin split, leading to a doubling of the conductance peaks when the excited states are doublets. We make here a qualitative comparison between the conductance measurements and the model in \eqref{eq:H} including the magnetotunneling Hamiltonian from  \eqref{Eq:H_MT}, for a different YSR state compared to main text. The comparative curves of the YSR state from the model and the measurements are shown in~\figrefadd{fig:figSfit}{}, Interestingly, we find that $\Gamma_\text{MT}$ has different amplitude and sign for different magnetic field directions, as expected for holes.
However, we emphasize this is not intended as a quantitative fit, but rather as a qualitative comparison between the induced asymmetry from the magnetotunneling term and the measured asymmetries. Indeed, as argued in~\ref{subsec:continuum}, the fitting parameters must be considered effective parameters rather than microscopic quantities due to the limitations of the model.

\newpage

\section{YSR\protect{s} at in-plane magnetic field}
\label{sec:sweeps}

\begin{figure*}[h!]
    \centering
    \includegraphics[width = \linewidth]{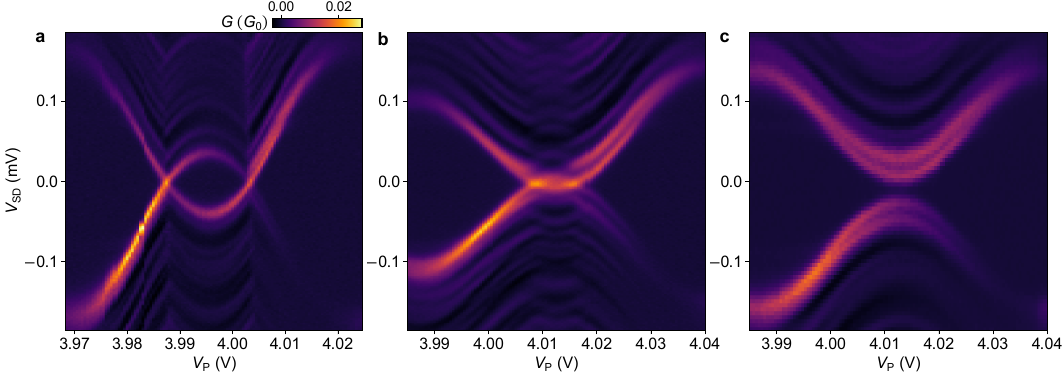}
    \caption{\textbf{YSRs in in-plane magnetic field.} \textbf{a-c.} $G$ vs $V_\mathrm{S}$ and $V_\mathrm{SD}$ at $B_x$ = 300 mT for the three configurations shown in Fig. 2c-e. In (a) and (b) we notice the same asymmetric splitting at the two sides of the doublet ground state region as observed in Fig. 3a-c, where an out-of-plane field of 80 mT was applied.}
    
    \label{fig:figS5}
\end{figure*}

\newpage

\section{$g$-tensor for intermediate coupling}
\label{sec:QPT g-tensor}

\begin{figure*}[h!]
    \centering
    \includegraphics[width = \linewidth]{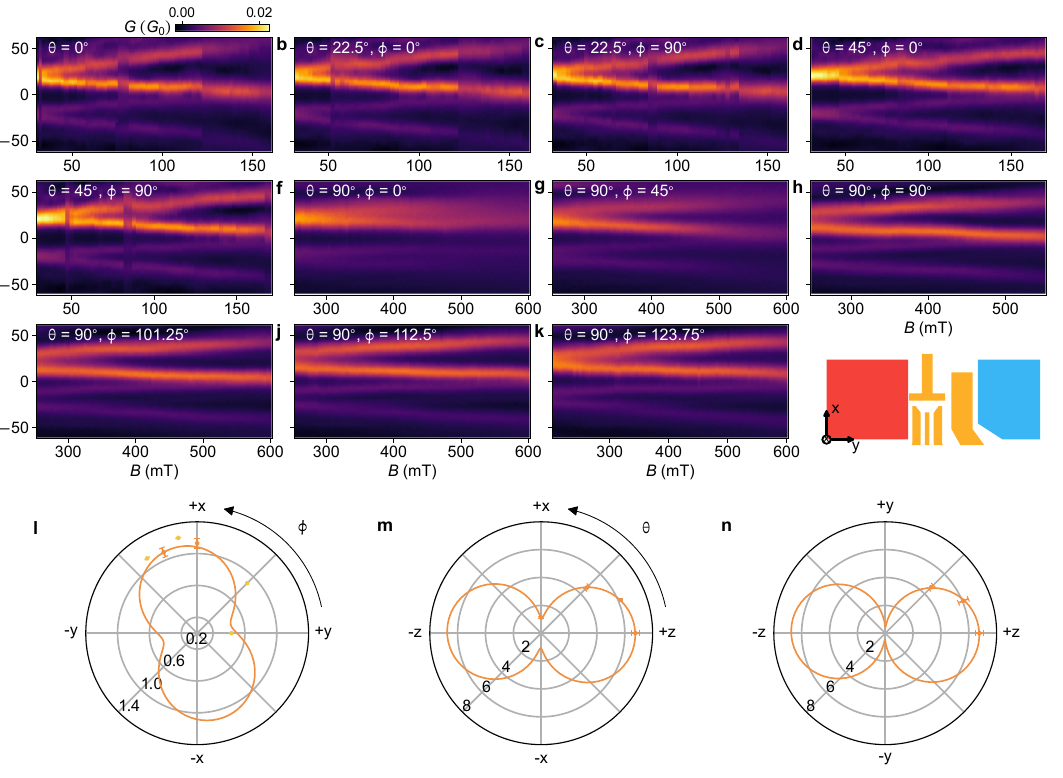}
    \caption{\textbf{$g$-tensor measurements for $\Gamma_\mathrm{S}$ = 183 $\mu$eV.} \textbf{a-k.} $G$ vs $V_\mathrm{SD}$ and magnetic field magnitude $B$ at different field directions indicated by the white labels in the top left of each plot. These are the raw data used to extract the $g$-tensor in l-m. \textbf{l-n.} Polar projections on the magnet axes of the $g$-tensor with the same scale as in Fig. 4.} 
   
    \label{fig:figS7}
\end{figure*}

\newpage

\section{Magnetic field sweeps raw data}
\label{sec:sweeps}

\begin{figure*}[h!]
    \centering
    \includegraphics[width = \linewidth]{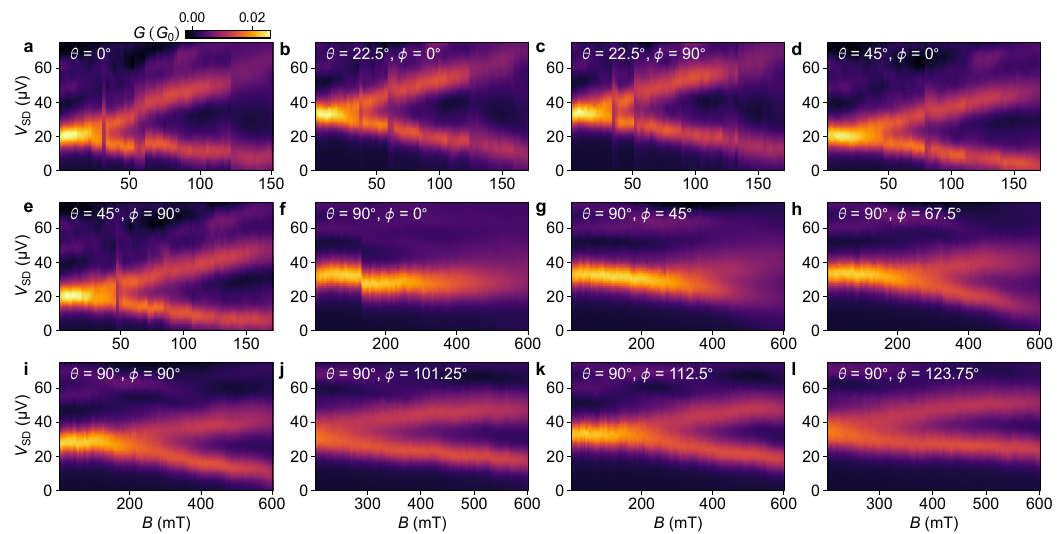}
    \caption{\textbf{Magnetic field sweeps for $\Gamma_\mathrm{S}$ = 156 $\mu$eV.} \textbf{a-l.} $G$ vs $V_\mathrm{SD}$ and magnetic field magnitude $B$ at different field directions indicated by the white labels in the top left of each plot. $\theta$ and $\phi$ are respectively the angle from the out-of-plane direction and the in-plane angle, as in main text. These are the raw data used to extract the $g$-tensor shown in Fig. 4a.}
    
    \label{fig:figS61}
\end{figure*}

\newpage

\begin{figure*}[!h]
    \centering
    \includegraphics[width = \linewidth]{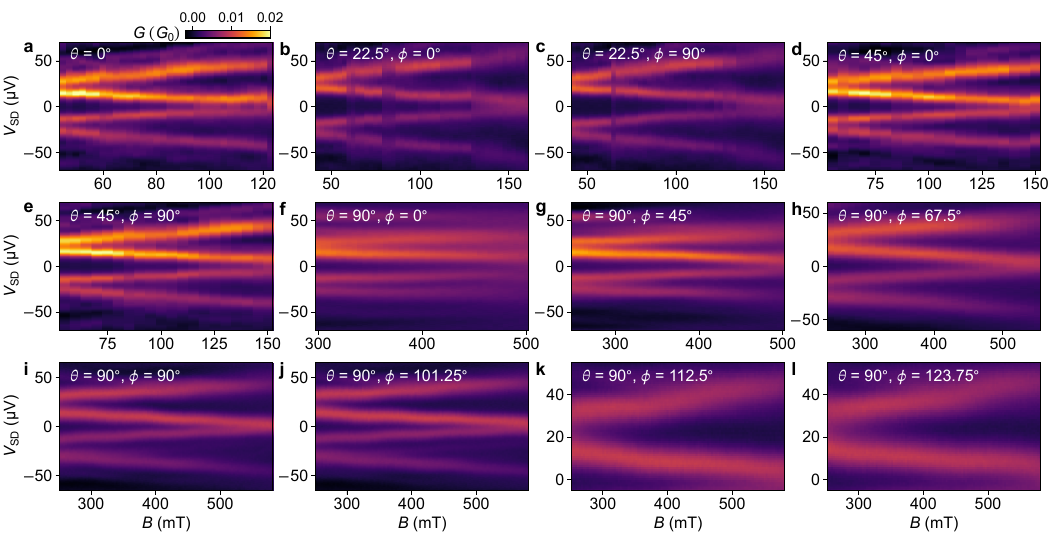}
    \caption{\textbf{Magnetic field sweeps for $\Gamma_\mathrm{S}$ = 204 $\mu$eV.} \textbf{a-l.} $G$ vs $V_\mathrm{SD}$ and magnetic field magnitude $B$ at different field directions indicated by the white labels in the top left of each plot. These are the raw data used to extract the $g$-tensor shown in Fig. 4b.}
    
    \label{fig:figS62}
\end{figure*}

\begin{figure*}[!h]
    \centering
    \includegraphics[width = \linewidth]{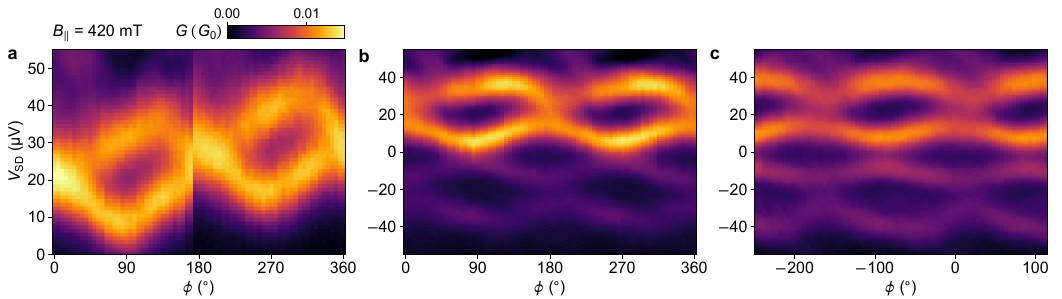}
    \caption{\textbf{In-plane magnetic field angle sweeps.} \textbf{a-c.} $G$ vs $V_\mathrm{SD}$ and in-plane magnetic field angle $\phi$ at $|B|$ = 420 mT. These are the raw data used to obtain the data shown in Fig. 4c. To extract the Zeeman splittings, a parabola fit on each of the two peaks was implemented to determine its position.}
    
    \label{fig:figS63}
\end{figure*}

\newpage

\section{$g$-tensor fitting procedure}
\label{sec:g-tensor fit}

For a spin-$\frac{1}{2}$ system characterized by a three-dimensional $g$-tensor $g$, the Zeeman splitting is given by \cite{CrippaPRL}:
\begin{equation}
    \Delta E=\mu_B \sqrt{\mathbf{B}^T g^T g \mathbf{B}} = \mu_B|\mathbf{B}|\sqrt{\mathbf{n}^TG\mathbf{n}} ~,
\end{equation}
where $\mathbf{B}$ is the applied magnetic field vector, $\mathbf{n}=\mathbf{B}/|\mathbf{B}|$ is the field direction and $G=g^Tg$ is a symmetric positive-definite matrix, square of the $g$-tensor. Experimentally, the effective $g$-factor in directions $\mathbf{n}_i$ is obtained from the linear dependence of the Zeeman splitting on $|\mathbf{B}_i|$, yielding:
\begin{equation}
    \label{eq:g}
    g_i = \sqrt{\mathbf{n}_i^TG\mathbf{n}_i}.
\end{equation}
Because $G$ is symmetric, it is fully specified by six independent parameters. Consequently, at least six measurements of $g_i$ along non-coplanar directions are required to determine the $g$-tensor.
To obtain the $g$-tensors we fit to equation \ref{eq:g} the $g$-factors measured along the nine directions marked by the dark-colored data points shown in Fig. 4a,b (raw data in \figrefadd{fig:figS61}{a-e,i-l} for $\Gamma_\mathrm{S} = 156~\mu\text{eV}$ and in \figrefadd{fig:figS62}{a-e,i-l} for $\Gamma_\mathrm{S} = 204~\mu\text{eV}$).

We point out that for $\theta \leq$ 45$^\circ$, the linear fits of $\Delta E$ versus $|\mathbf{B}|$ extrapolate to finite intercepts on the field axis, ranging from 9 to 17 mT. Although we did not investigate this effect in detail, we speculate that it may arise from vortex formation in the superconductor. At $\theta =$ 90$^\circ$ (in-plane), we restricted the fit to those directions where the extrapolated crossing points lie within $\pm$20 mT. For $\phi = 0, 45^\circ ~\text{and}~ 67.5^\circ$ the linear fit extrapolates to $\Delta E =0$ at $|B|>80$ mT (raw data in \figrefadd{fig:figS61}{f-h} for $\Gamma_\mathrm{S} = 156~\mu\text{eV}$ and in \figrefadd{fig:figS62}{f-h} for $\Gamma_\mathrm{S} = 204~\mu\text{eV}$). Here, instead of relying on the measured slope, we use the absolute Zeeman splitting measured at $|B| = 350~\text{mT}$ and compare it to the value obtained at $\phi = 90^\circ$, resulting in the light-colored data points. However, we stress that these three directions were not used to fit the $g$-tensor. 

For the configuration with $\Gamma_S = 156 \mu\text{V}$ (Fig. 4a), the direction $\theta = 22.5^\circ$, $\phi=90^\circ$ (marked by the light-blue triangle in Fig. 4a) was also excluded from the fit. Including it leads to a $g$-tensor whose in-plane projection does not reproduce our data points. 

Finally, in the fit for $\Gamma_S = 204 \mu\text{V}$ (Fig 4b) we impose a lower bound on the $g$-factor at $\theta = 90^\circ$, $\phi = 0^\circ$ (y-direction). Without this constraint, the fit yields $g_y=0.45$, which would correspond to a Zeeman splitting of 9 $\mu$V at 350 mT, too small to produce two resolvable peaks. Experimentally, however, we clearly resolve two peaks at this field (\figrefadd{fig:figS62}{f}), with a splitting of 13 $\mu$V, corresponding to $g_y = 0.64$. We therefore restrict the fit to satisfy $g_y \geq 0.64$.\\ 
Following this procedure we obtain the $g$-tensors shown in Fig. 4 of main text and Fig. \ref{fig:figS7}. Here we report the tensors in matrix form for the three coupling configurations: 

\begin{equation}
    G_{\Gamma_\mathrm{S}=156~\mu\text{eV}} = 
    \begin{pmatrix} 
    0.76 & -0.18 & 0.65\\ 
    -0.18 & 0.09 & 1.48\\
    0.65 & 1.48 & 55.40\\
    \end{pmatrix}
    \longrightarrow
    \underline{g}_{156~\mu\text{eV}} = 
    \begin{pmatrix} 
    0.90 & 0 & 0\\ 
    0 & \sim0 & 0\\
    0 & 0 & 7.45\\
    \end{pmatrix},
\end{equation}
\begin{equation}
    G_{\Gamma_\mathrm{S}=183~\mu\text{eV}} = 
    \begin{pmatrix} 
    1.17 & -0.24 & -1.56\\ 
    -0.24 & 0.25 & -1.76\\
    -1.56 & -1.76 & 45.54\\
    \end{pmatrix}
    \longrightarrow
    \underline{g}_{183~\mu\text{eV}} = 
    \begin{pmatrix} 
    1.10 & 0 & 0\\ 
    0 & 0.30 & 0\\
    0 & 0 & 6.76\\
    \end{pmatrix},
\end{equation}
\begin{equation}
    G_{\Gamma_\mathrm{S}=204~\mu\text{eV}} = 
    \begin{pmatrix} 
    1.42 & -0.09 & -2.74\\ 
    -0.09 & 0.42 & -2.62\\
    -2.74 & -2.62 & 44.65\\
    \end{pmatrix}
    \longrightarrow
    \underline{g}_{204~\mu\text{eV}} = 
    \begin{pmatrix} 
    1.14 & 0 & 0\\ 
    0 & 0.46 & 0\\
    0 & 0 & 6.71\\
    \end{pmatrix}.
\end{equation}

where the $G$ matrices are written in the magnet-axes basis $(x, y, z)$. We diagonalize $G$, so that $G_d=WGW^\dagger$ is expressed in the principal-axes basis. Finally, we define $\underline{g}=G_d^{1/2}$ which expresses the effective $g$-matrix in the principal axes basis.

The Zeeman-splitting measurements determine $G=g^Tg$, but not a unique signed $g$-matrix. For the tensor decomposition and the control estimates of \secref{sec:sm_tensor_extraction} we therefore choose the symmetric positive-semidefinite representative in the \emph{laboratory} basis, $g_\alpha=G_\alpha^{1/2}$, where $\alpha$ can be any of the measured configurations. This is one admissible gauge: rotating $g$ with any orthogonal matrix in spin space, gives the same measured $G$. We use the same symmetric representative at the three gate settings, thereby retaining the $g$-matrix-modulation contribution inferred from the static tensors, but not an independent gate-dependent spin-space rotation~\cite{CrippaPRL}.

\newpage
\section{Suppressed Zeeman splitting}
\label{sec:zerosplitting}

\begin{figure*}[htbp!]
    \centering
    \includegraphics[width = \linewidth]{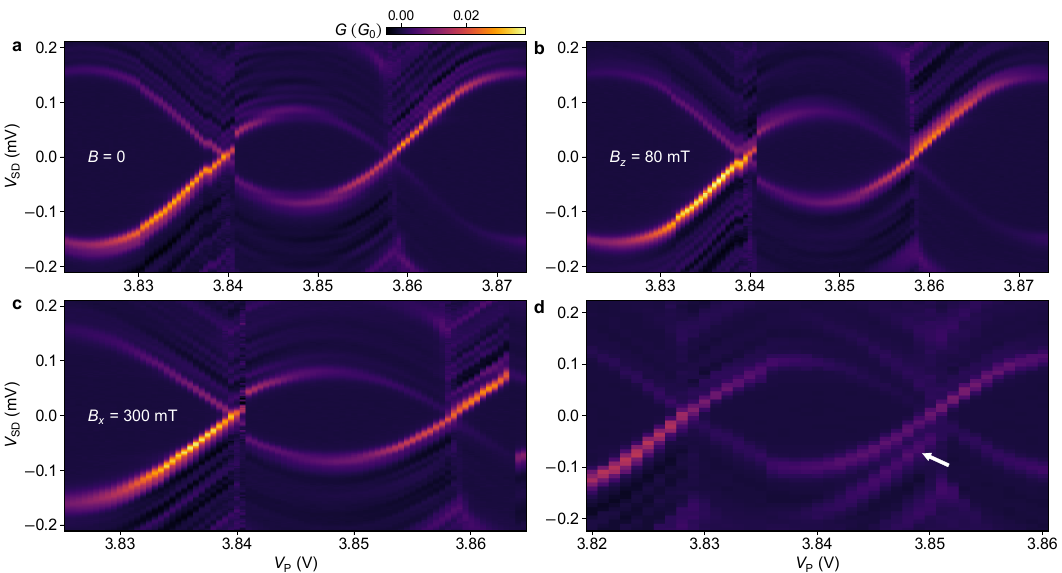}
    \caption{\textbf{YSRs with suppressed Zeeman splitting.} \textbf{a-c.} $G$ vs $V_\mathrm{SD}$ and $V_\mathrm{P}$ at zero magnetic field (a), 80 mT out-of-plane (b) and 300 mT in-plane (along $x$-direction), measured for the same YSR state studied in the main text, but at a barrier voltage $V_\mathrm{S}$ that is 180 mV higher. No Zeeman splitting is observed. In panel a we note a nonconventional behavior of the YSR feature in the center of the doublet ground state region. \textbf{d.} Same YSR state as in a-c, measured at zero magnetic field and at a $V_S$ that is 30 mV higher. We highlight the appearance of a close-by second orbital, indicated by the white arrow.}
    
    \label{fig:figS8}
\end{figure*}

\newpage
\section{Magnetic field misalignment}
\label{sec:zerosplitting}

\begin{figure*}[htbp!]
    \centering
    \includegraphics[width = \linewidth]{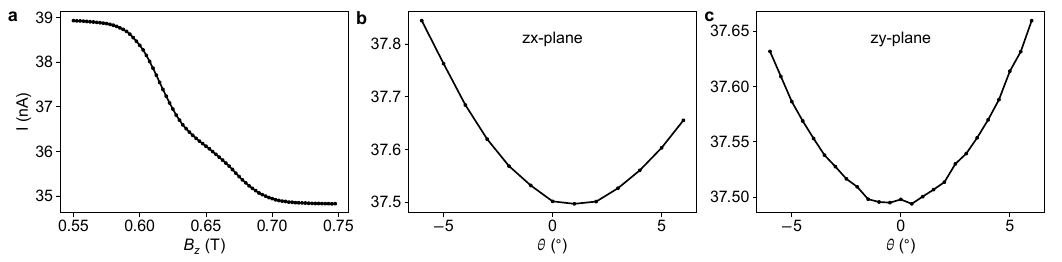}
    \caption{\textbf{Magnetic field misalignment measurement.} \textbf{a.} Current $I$ measured across a grAl stripe as function of $B_z$, using a two-probe geometry.
    We set a bias voltage of 1 mV over the 25 k$\Omega$ line resistance, resulting in a 40 nA current, below the grAl critical current. As $B_z$ is increased, grAl transitions to the normal state, leading to an increase in the total resistance and a corresponding decrease in the measured current. \textbf{b-c.} I as a function of the angle from the out-of-plane direction $\theta$ in the zx-plane (b) and zy-plane (c) at $|B|$ = 615 mT, corresponding to the maximum slope of panel a. A parabolic fit of the zx-plane (zy) trace results in a measured misalignment of 1.2$^\circ$ (-0.3$^\circ$).}
    
    \label{fig:figS9}
\end{figure*}

\newpage

\section{Tensor extraction and hybridization-driven \texorpdfstring{$g$}{g}-matrix modulation}
\label{sec:sm_tensor_extraction}
The ZBW model of \secref{sec:magnetotunnel} is written for a single field direction, in terms of scalar effective $g$-factors. Here we extend it to signed $g$-matrices, so that the three tensors measured in \secref{sec:g-tensor fit} can be combined into a single data-informed description, and we use it to estimate the spin rotation produced by a modulation of the QD-superconductor coupling $\GS$. Throughout, $G=g^Tg$ denotes the measured symmetric tensor, $g$ a $g$-matrix, and $\gMT$ the tunnel $g$-matrix. The scalar convention of \eqref{Eq:H_MT} is generalized to
\begin{equation}
    H_\mathrm{T}=d^\dagger\left[\GS\mathbb{I}-\frac{\mu_B}{2}\boldsymbol{\sigma}\cdot \gMT(\GS)\mathbf{B}\right]c+\mathrm{H.c.},
    \label{eq:HT_tensor_sm}
\end{equation}
which reduces to \eqref{Eq:H_MT} with $\Gamma_\mathrm{MT}=\frac{\mu_Bg_{\mathrm{MT}}B}{2}$ in the collinear limit.

The three tensors were measured at $(\GS,n_g)=(156~\mu\text{eV},0.58)$, $(183~\mu\text{eV},0.63)$ and $(204~\mu\text{eV},1)$; the first two positions are quoted as $\Delta V_\mathrm{P}=1.42$ and $1.37$ in the coordinate of Fig.~3 of the main text and are converted using $n_g=2-\Delta V_\mathrm{P}$, placing both in region~II. Their laboratory-frame symmetric square roots $g_{156}$, $g_{183}$ and $g_{204}$ follow from  the measured tensors.

Projecting the QD, superconductor and tunnel Zeeman terms onto the lowest odd doublet of the zero-field block \eqref{eq:Hodd} (i.e.\ at $\Gamma_\text{MT}=0$ and $B=0$), with normalized eigenvector $u=(u_1,u_2,u_3,u_4)^T$ in the basis $\mathcal{B}_{+}=\{\ket{\uparrow\downarrow_{QD},\uparrow_{SC}},\ket{\uparrow_{QD},\uparrow\downarrow_{SC}},\ket{\uparrow_{QD},0_{SC}},\ket{0_{QD},\uparrow_{SC}}\}$, gives
\begin{equation}
    \geff=w_\mathrm{QD}\gQD+w_\mathrm{SC}\gSC-w_T\,\gMT(\GS),
    \label{eq:geff_sm}
\end{equation}
with the QD, superconductor and coherent tunnel weights
\begin{equation}
    w_\mathrm{QD}=|u_2|^2+|u_3|^2,\quad
    w_\mathrm{SC}=|u_1|^2+|u_4|^2=1-w_\mathrm{QD},\quad
    w_T=2\,\mathrm{Re}\left(u_1^*u_2+u_3^*u_4\right).
\end{equation}

Because $w_T$ vanishes at $n_g=1$, the $\GS=204~\mu$eV tensor fixes the QD matrix once the superconducting one is set to the signed hole convention $\gSC=-2\mathbb{I}$ (scalar $g$-factor with relative negative sign), 
\begin{equation}
    \gQD=\frac{g_{204}-w_{\mathrm{SC},204}\,\gSC}{w_{\mathrm{QD},204}}
    =\begin{pmatrix}
    -1.48 & -0.15 & -0.40\\
    -0.15 & -0.79 & -0.41\\
    -0.40 & -0.41 & -7.63
    \end{pmatrix},
\end{equation}
with principal magnitudes $(0.73,\,1.50,\,7.68)$, and the region-II point at $\GS=156~\mu$eV then fixes the tunnel matrix,
\begin{equation}
    \gMT^{156}=-\frac{g_{156}-w_{\mathrm{QD},156}\,\gQD-w_{\mathrm{SC},156}\,\gSC}{w_{T,156}}
    =\begin{pmatrix}
    -0.46 & -0.24 & -1.20\\
    -0.24 & -0.98 & -1.56\\
    -1.20 & -1.56 & -3.12
    \end{pmatrix}.
\end{equation}
Following the argument of \secref{sec:magnetotunnel} that the magnetotunneling amplitude scales with the kinetic overlap generating the bare tunnel coupling, we take
\begin{equation}
    \gMT(\GS)=\frac{\GS}{156~\mu\text{eV}}\,\gMT^{156}.
    \label{eq:gMT_scaling}
\end{equation}
These are effective matrices inferred under the assumptions $\gSC=-2\mathbb{I}$, the gate-independent $g$-matrices $\gQD\,,\gSC\,,\gMT$; they are not directly measured bare parameters, and the individual entries should not be over-interpreted. Within the chosen gauge $\gMT$ is dominated by its out-of-plane entry, as expected from the predominantly heavy-hole character of the states and consistent with the strongly anisotropic spin-selective tunneling of Ref.~\cite{katsaros2011spinselective}, but it is far from purely longitudinal. The tunnel matrix is neither scalar nor proportional to $\gQD$, in line with the $g$-matrix formulation of magnetotunneling of Ref.~\cite{rodriguez2025sweet}. This misalignment is what makes the drive of Fig.~5(a-c) effective: $\gQD\mathbf{B}$ and $\gMT\mathbf{B}$ compete for the orientation of the Larmor vector along $\hat{x}$, $\hat{y}$ and $\hat{z}$, whereas $\gMT\propto\gQD$ would render the $\GS$ modulation purely longitudinal and $f_\mathrm{R}$ would vanish.

We note the intermediate $\GS=183~\mu$eV tensor, which was not used in the calibration, is reproduced only approximately, with relative Frobenius errors of $18\%$ on $g_{183}$. This residual quantifies the limitations of treating $\gQD$, $\gSC$ and the magnetotunneling scaling as gate-independent effective quantities across all three electrostatic configurations, and we therefore use the extraction as a data-informed estimate rather than as a unique microscopic decomposition.

At fixed $n_g$, both the weights and the tunnel matrix depend on $\GS$, so that
\begin{equation}
    \frac{\partial\geff}{\partial\GS}
    =\frac{\partial w_\mathrm{QD}}{\partial\GS}\gQD
    +\frac{\partial w_\mathrm{SC}}{\partial\GS}\gSC
    -\frac{\partial w_T}{\partial\GS}\gMT(\GS)
    -\frac{w_T}{156~\mu\text{eV}}\gMT^{156},
    \label{eq:dgdGamma}
\end{equation}
the last term originating from the explicit scaling \eqref{eq:gMT_scaling}. For a weak harmonic modulation $\GS(t)=\GS^0+\delta\GS\sin(2\pi ft)$ of peak amplitude $\delta\GS$, the static and modulated Larmor vectors are $\mu_B\geff\mathbf{B}$ and $\mu_B\geff'\mathbf{B}\,\delta\GS$, with $\geff'\equiv\partial\geff/\partial\GS$, and only the component of the latter transverse to the former drives rotations. The $g$-tensor-modulation ($g$-TMR) contribution to the Rabi frequency is therefore
\begin{equation}
    \frac{f_\mathrm{R}}{\delta\GS}
    =\frac{\mu_B}{2h}\,
    \frac{\left|\left(\geff\mathbf{B}\right)\times\left(\geff'\,\mathbf{B}\right)\right|}{\left|\geff\mathbf{B}\right|},
    \label{eq:fR_sm}
\end{equation}
which is linear in $|\mathbf{B}|$ and in the drive amplitude. Parametrizing the field as $\mathbf{B}=B(\sin\theta\sin\phi,\,\sin\theta\cos\phi,\,\cos\theta)^T$, \eqref{eq:fR_sm} evaluated at $\GS^0=156~\mu$eV, $n_g=0.58$ and $|\mathbf{B}|=100$~mT gives the map of Fig.~5a, with a maximum of $7.9~\mathrm{MHz}/\mu$eV near $(\theta,\phi)\simeq(42^\circ,210^\circ)$. 

Finally, we remark that a static splitting measurement determines $G=g^Tg$, while the $g$-matrix is defined only up to a left orthogonal rotation in spin space. The symmetric square-root choice provides a continuous representative and hence a definite $g$-matrix-modulation estimate, but a gate-dependent rotation would generate an additional iso-Zeeman contribution that the present static spectroscopy cannot extract. The rates in Fig.~5(a-c) should therefore be read as the data-informed $g$-TMR contribution within the chosen gauge, not as the total electrically driven Rabi frequency.

\newpage
\section{Phase-dependent spin curvature}
\label{sec:sm_two_sc_curvature}

We now consider a minimal extension in which the QD is contacted by two superconducting orbitals with phase difference $\phiLR$. This is the simplest setting in which the spin-dependent tunneling identified above becomes phase sensitive, and therefore resonator detectable. Each lead $\alpha=L,R$ carries a pairing term with phase $\varphi_\alpha=\pm\phiLR/2$, a Zeeman term with matrix $g_{\mathrm{SC},\alpha}$, and a scalar tunnel amplitude $\Gamma_\alpha$ dressed by the magnetotunneling correction of \eqref{eq:HT_tensor_sm},
\begin{equation}
    H_{dS}=\sum_{\alpha=L,R}\left[d^\dagger\left(\Gamma_\alpha\mathbb{I}-\frac{\mu_B}{2}\boldsymbol{\sigma}\cdot g_{\mathrm{MT},\alpha}\mathbf{B}\right)c_\alpha+\mathrm{H.c.}\right],
\end{equation}
while the two leads are coupled by a coherent cotunneling $t_c\sum_\sigma(c^\dagger_{L\sigma}c_{R\sigma}+\mathrm{H.c.})$--assumed spin-conserving for simplicity. The retained model thus contains no independent zero-field spin-flip tunnel amplitude, no Peierls phase and no magnetocotunneling.

The local, one-lead magnetotunneling correction to the effective ground-state $g$-matrix is perturbatively inversely proportional to the energy cost of adding or removing a particle from the QD $\lambda_2^{-1}-\lambda_0^{-1}$, where $\lambda_0=\Delta+U(n_g-\frac12)$ and $\lambda_2=\Delta+U(\frac32-n_g)$ are the energies of the virtual excursions to the empty and doubly occupied dot. It therefore cancels at $n_g=1$, where $\lambda_0=\lambda_2\equiv\lambda=\Delta+U/2$, consistently with $w_T=0$ in \secref{sec:sm_tensor_extraction}. We work at this point, so that any residual spin dependence of the tunneling is necessarily of crossed origin: paths $d\leftrightarrow L\leftrightarrow R\leftrightarrow d$ in which one link supplies a scalar and the other a spin-dependent component from magnetotunneling, with matrix structure $\Gamma_R g_{\mathrm{MT},L}+\Gamma_L g_{\mathrm{MT},R}$. To lowest order they generate a phase-dependent effective matrix $\geff(\phiLR)=g_0+g_\varphi\cos\phiLR$ which, in the identical-link limit $\Gamma_L=\Gamma_R=\GS$, $g_{\mathrm{SC},\alpha}=\gSC$ and $g_{\mathrm{MT},\alpha}=\gMT$, reads
\begin{equation}
\begin{aligned}
    g_0={}&\left(1-\frac{2\GS^2}{\lambda^2}\right)\gQD+\frac{2\GS^2}{\lambda^2}\gSC-2\GS t_c\left(\frac{1}{\lambda^2}-\frac{1}{\Delta\lambda}\right)\gMT,\\
    g_\varphi={}&2\GS t_c\left(\frac{1}{\lambda^2}+\frac{1}{\Delta\lambda}\right)\gMT.
\end{aligned}
\label{eq:g0gphi}
\end{equation}
This mechanism does not require an additional zero-field spin-flip tunneling amplitude within the retained model: the phase sensitivity of the spin splitting is inherited entirely from the field-linear magnetotunneling terms.

The two Zeeman branches of the odd doublet are $E_s(\phiLR)=E_D(\phiLR)+s\frac{\mu_B}{2}|\geff(\phiLR)\mathbf{B}|$ with $s=\pm1$, where the scalar part
\begin{equation}
    E_D(\phiLR)=-2\Delta-\frac{\Gamma_L^2+\Gamma_R^2}{\lambda}-\frac{t_c^2}{2\Delta}\left(1+\cos\phiLR\right)
    \label{eq:ED}
\end{equation}
carries the Josephson phase dependence.

We now consider the coupling to a microwave cavity which is represented as a lumped-element LC resonator with bare resonance frequency $f_{\mathrm{res}}=\frac{1}{2\pi\sqrt{LC}}$. Using photon operators, the resonator can be modeled as $\hat H=f_{\mathrm{res}}a^\dagger a$. Assuming a negligible loop inductance, the dc phase difference across the QD $\phiLR = 2\pi\Phi/\Phi_0$ is imposed by the magnetic flux threading the superconducting loop $\Phi$ in units of the flux quantum $\Phi_0 = h/2e$ (see Fig. 5e in the main text). Current fluctuations in the resonator couple to the QD junction, which induce phase fluctuations $\delta\hat\varphi_\mathrm{LR}=\phiLR+\hat\varphi$, with $\hat\varphi=\varphi_\mathrm{ZPF}(a+a^\dagger)$ being the phase operator of the resonator characterised by the zero-point phase fluctuations $\varphi_\mathrm{ZPF}=\sqrt{\pi Z/R_Q}$, with amplitude given by the ratio between the resonator impedance $Z=\sqrt{L/C}$ and the superconducting quantum of resistance $R_Q=h/4e^2$. By
expanding the QD junction Hamiltonian up to second order in $\delta\hat\varphi$, the adiabatic inductive shift of the resonator is ~\cite{park2020adiabatic}
\begin{equation}
\label{inductive-shift}
\frac{h\delta f_{\mathrm{res},s=\pm 1}}{\varphi_\mathrm{ZPF}^2}=\,\partial^2E_s/\partial\phiLR^2=\left[\frac{t_c}{2\Delta}\mp\mu_B\left(\frac{1}{\lambda^2}+\frac{1}{\Delta\lambda}\right)\gMT\right]t_c\cos\phiLR.
\end{equation}
Physically, this shift can be interpreted as a renormalization of the resonator inductance by the effective QD inductance $L_s=(\Phi_0/2\pi)^2 (\partial^2E_s/\partial\phiLR^2)^{-1}$ which, importantly, is spin-dependent ($s=\pm 1$). Further qubit properties, including the full dynamical susceptibility response beyond this dispersive shift, will be discussed elsewhere~\cite{pino2026}.

Figure~5(d,e) in the main text  plots the resonator shift in \eqref{inductive-shift} 
for the two spins $s=+1$ and $s=-1$ (labelled state~0 and state~1). Since \eqref{eq:ED} contributes a common shift to both branches, the differential response $\Delta f_\mathrm{res}=\delta f_{\mathrm{res},1}-\delta f_{\mathrm{res},0}$ (black curve) is the readout-relevant signal; because $|\geff\mathbf{B}|$ is not a pure harmonic of $\phiLR$, it generally contains higher harmonics and a phase-independent offset, most visible for in-plane fields where $|g_0\mathbf{B}|$ is small.

The curves are computed with $U=685~\mu$eV, $\Delta=200~\mu$eV, $\Gamma_L=\Gamma_R=156~\mu$eV, $t_c=16~\mu$eV and $n_g=1$ (so $\lambda=542.5~\mu$eV), at $|\mathbf{B}|=100$~mT along $\hat{z}$ (panel~d) and $\hat{x}$ (panel~e), using the $\gQD$, $\gSC=-2\mathbb{I}$ and $\gMT(\GS)$ of \secref{sec:sm_tensor_extraction}. We notice $t_c$ is an effective parameter of the hypothetical two-terminal geometry and is not extracted from the present single-superconductor experiment, so the absolute curvature magnitude is illustrative: from \eqref{eq:g0gphi} and \eqref{eq:ED}, the common Josephson contribution scales as $t_c^2/\Delta$, while the spin-dependent part scales linearly with $t_c$, with $\GS$ and with $|\mathbf{B}|$.

\balancecolsandclearpage

\end{document}